\documentclass[twocolumn,tighten]{aastex62}
\usepackage{lineno}
\usepackage{natbib}

\newcommand{\apg}  	{^{>}_{\sim}}
\newcommand{\apll}  	{^{<}_{\sim}}

\newcommand{\hst}	{{\it HST}}

\newcommand{\farcc}{\mbox{\ensuremath{,\!\!^{\prime\prime}}}}

\received{May 11, 2018}
\accepted{November 9, 2018}
\submitjournal{ApJ}

\shorttitle{Strongly-lensed DSFGs}
\shortauthors{Frye et al.}

\usepackage{graphicx}

\begin{document}

\title{\bf PLCK G165.7+67.0:  Analysis of a Massive Lensing Cluster  in a Hubble Space Telescope Census of Submillimeter  Giant Arcs Selected Using {\it Planck/Herschel}}

\correspondingauthor{Brenda L.~Frye}
\email{bfrye@as.arizona.edu}

\author{Brenda L.~Frye}
\affil{Department of Astronomy/Steward Observatory, 933 North Cherry Avenue, University of Arizona, Tucson, AZ  85721,  USA; bfrye@as.arizona.edu}

\author{Massimo Pascale}
\affil{Department of Astronomy/Steward Observatory, 933 North Cherry Avenue, University of Arizona, Tucson, AZ  85721, USA; bfrye@as.arizona.edu}

\author{Yujing Qin}
\affil{Department of Astronomy/Steward Observatory, 933 North Cherry Avenue, University of Arizona, Tucson, AZ  85721,  USA; bfrye@as.arizona.edu}

\author{Adi Zitrin}
\affiliation{Physics Department, Ben-Gurion University of the Negev, P.~O.~Box 653, Be'er-Sheva, 8410501,  Israel}

\author{Jos{\'e} Diego}
\affiliation{IFCA, Instituto de Fíisica de Cantabria (UC-CSIC), Av. de Los Castros s/n, E-39005 Santander, Spain}

\author{Greg Walth}
\affiliation{Center for Astrophysics \& Space Sciences, University of California at 
San Diego, 9500 Gilman Drive, La Jolla, CA 92093-0424, USA}

\author{Haojing Yan}
\affiliation{Department of Physics and Astronomy, University of Missouri, Columbia, MO 65211, USA}

\author{Christopher J.~Conselice}
\affiliation{School of Physics and Astronomy, The University of Nottingham, University Park, Nottingham, NG7 2RD, UK}

\author{Mehmet Alpaslan}
\affiliation{Center for Cosmology and Particle Physics, New York University, 726 Broadway, New York, NY 10003,  USA}

\author{Adam Bauer}
\affiliation{Department of Astronomy/Steward Observatory, 933 North Cherry Avenue, University of Arizona, Tucson, AZ  85721,  USA}

\author{Lorenzo Busoni}
\affiliation{Osservatorio Astrofisico di Arcetri, Largo Enrico Fermi 5, I-50125, Florence, Italy}

\author{Dan Coe}
\affiliation{STScI, 3700 San Martin Drive, Baltimore, MD  21218, USA}

\author{Seth H.~Cohen}
\affiliation{School of Earth \& Space Exploration, Arizona State University, Tempe, AZ 85287-1404, USA}

\author{Herv{\'e} Dole}
\affiliation{Institut d'Astrophysique Spatiale, CNRS, Univ.~Paris-Sud, Universit\'e Paris-Saclay, F-91400 Orsay, France}

\author{Megan Donahue}
\affiliation{Physics \& Astronomy Department, Michigan State University, East Lansing, MI 48824-2320, USA}

\author{Iskren Georgiev}
\affiliation{Max Planck Institut f{\"u}r Astronomie, K{\"o}nigstuhl 17, D-69117, Heidelberg, Germany}

\author{Rolf A.~Jansen}
\affiliation{School of Earth \& Space Exploration, Arizona State University, Tempe, AZ 85287-1404, USA}

\author{Marceau Limousin}
\affiliation{Aix Marseille Universit{\'e}, LAM, Laboratoire d'Astrophysique de Marseille, Marseille, France}

\author{Rachael Livermore}
\affiliation{School of Physics, University of Melbourne, Melbourne, Victoria, Australia}

\author{Dara Norman}
\affiliation{National Optical Astronomical Observatory, 950 N.~Cherry Ave., Tucson, AZ  85719, USA}

\author{Sebastian Rabien}
\affiliation{Max Planck Institut f{\"u}r Extraterrestrische Physik, Garching, Germany}

\author{Rogier A.~Windhorst}
\affiliation{School of Earth \& Space Exploration, Arizona State University, Tempe, AZ 85287-1404, USA}

\begin{abstract}
We present {\it Hubble Space Telescope} WFC3-IR imaging in the fields of six apparently bright 
dusty star-forming galaxies (DSFGs) at $z$\,=\,2\,--\,4 identified by their rest-frame 
far-infrared colors using the {\it Planck} and {\it Herschel} space facilities.  We 
detect near-infrared counterparts for all six submillimeter sources, allowing us to
undertake strong-lensing analyses.  One field in particular stands out for its prominent 
giant arcs, PLCK G165.7+67.0 (G165).  After combining the color and morphological 
information, we identify 11 sets of image multiplicities in this one field.  We 
construct a strong-lensing model constrained by this lensing evidence, which 
uncovers a bimodal spatial mass distribution, and from which we measure a 
mass of $(2.6 \pm 0.11)$\,$\times$\,$10^{14}$\,$M_{\odot}$  within $\sim$250 kpc.  
The bright ($S_{350}$\,$\approx$\,750 mJy) DSFG appears as two images:  a giant 
arc with a spatial extent of 4$\farcs$5 that is merging with the critical curve, and a 
lower-magnification counterimage that is detected in our new longer-wavelength 
ground- and space-based imaging data.  Using our ground-based spectroscopy, we 
calculate a dynamical mass of $1.3^{+0.04}_{-0.70} \times 10^{15}$\,$M_{\odot}$ to 
the same fixed radius, although this value may be inflated relative to the true value if 
the velocity distribution is enhanced in the line-of-sight direction.  We suggest that the 
bimodal mass taken in combination with the weak  X-ray flux and low SZ decrement 
may be explained as a pre-merger for which the intracluster gas is diluted along the 
line of sight, while the integrated surface mass density is supercritical to strong-lensing 
effects.

\end{abstract}

\keywords{gravitational lensing: strong --  galaxies:  fundamental parameters -- galaxies:  clusters:  general --galaxies:  high-redshift --submillimeter:  galaxies}


\section{Introduction}

Clusters of galaxies with masses $\sim$$10^{15}$\,M$_{\odot}$ are extremely useful but 
rare tracers of the distribution of mass in the universe \citep{Bahcall:77, Mo:96}.  Finding  
galaxy clusters, and then establishing their cluster properties and cluster scaling relations are 
fundamental to cosmology studies \citep{Vikhlinin:09, Mantz:10, Rozo:10, Allen:11, Benson:13, 
Hasselfield:13, Planck:14}.   As ensembles of discrete galaxies, clusters can be discovered 
in optical and near-infrared (NIR) wide-area surveys{\bf ,} such as the Sloan Digital Sky Survey 
\citep[SDSS; i.e.,][]{Koester:07a, Koester:07b,Rykoff:14,Rykoff:16}. 

Although originally discovered at optical wavelengths, galaxy clusters with masses of 
(1\,--\,10)\,$\times$\,10$^{15}$\,M$_{\odot}$ will almost always contain a massive 
component of hot intracluster gas, which makes them distinct X-ray sources.  This 
reservoir of hot baryons is a salient feature of massive clusters, as there is no physical 
mechanism to dissipate it.  To search for this requisite feature, the ROSAT archives 
offer the all-sky advantage to efficiently detect the most extreme sources of X-ray emission 
produced by thermal bremsstrahlung and line emission in the intracluster gas 
\citep{Rosati:98, Ebeling:07, Ebeling:10}.

A galaxy  cluster bound by gravity also has a distinct signature at radio wavelengths.  
This is because the same large reservoirs of intracluster gas that give rise to the 
X-ray flux also distort the cosmic microwave background  (CMB) radiation by inverse 
Compton scattering.  From the ground, searches for galaxy clusters by the detection of 
this Sunyaev-Zel'dovich (SZ) effect using the South Pole Telescope 
\citep[SPT;][]{Carlstrom:11} yield hundreds of candidates \citep{Bleem:15}.  Targeted 
searches using the Atacama Cosmology Telescope \citep[ACT; ][]{Fowler:07} that 
exercise a similar approach are also successful \citep{Sehgal:11,Sehgal:13}.  From 
space, {\it Planck} High Frequency Imager (HFI)  data {\bf are} used to extend the search 
for the SZ decrement to all available  extragalactic sky \citep{Lamarre:03, PlanckXXVII:16}.

To complement the {\it cosmological} SZ approach of searching for clusters, the detection of 
apparently bright galaxies by the {\it astronomical} technique of color selection 
has recently been explored.  For example, the infrared-bright dusty star forming 
galaxies (DSFGs) produce stars at rates of up to $\sim$1000 $M_{\odot}$\,yr$^{-1}$ 
and yield prodigious amounts of dust.  This warm dust radiates as a modified blackbody 
spectrum with a prominent
peak in the rest-frame far-infrared.   Submillimeter data are well suited to conduct 
the color search for the DSFGs{\bf ,} because this wavelength range corresponds
to the observed-frame thermal dust peak at redshifts typical of DSFGs of 
$z$\,$\approx$\,2\,--\,4  \citep[][and references therein]{Casey:14, Planck:15}.  
In this regime, there is the unusual advantage that one records the flux density 
of the DSFGs closer to the peak of their rest-frame spectral energy distributions (SEDs)
 as their redshift increases.  
As a result, the benefit of the high flux density of DSFGs largely compensates for 
the cosmological dimming \citep{Blain:99b, Planck:15}, thereby gaining leverage for 
the detection of high-redshift objects.

A {\it Planck}/HFI \citep{PlanckXXVII:16} census was undertaken to find DSFGs by color 
covering the portion of the sky with minimal cirrus contamination, which we take
to mean that the column density of hydrogen in the 857 GHz map is less than the 
minimum value of 
$N_{HI}$\,$=$\,3\,$\times$\,$10^{20}$ cm$^{-1}$, amounting to 26\% of the sky.
To be
selected by {\it Planck}/HFI, the DSFGs had to be separately detected
in each of the cleaned 857, 545 and 353~GHz maps, be compact at {\it
  Planck} resolution  ($\sim$4.5$^{\prime}$), and have flux density
ratios in the 353, 545, and 857 GHz maps consistent with being red and
dusty sources \citep{Planck:15}.  These so-called `cold' sources of
the cosmic infrared background \citep[CIB][]{Puget:96,Dole:06}
are extremely rare at the {\it Planck} point-source sensitivity
of about $\sim$\,600 mJy at $\sim$\,545 GHz, with number densities of $\sim$1 per
few square degrees, requiring the wide-field survey area of {\it
  Planck}/HFI. \cite{Planck:16} selected this way $\sim$2000 DSFG candidates.

To classify the sources, \cite{Planck:15} performed follow-up
observations at higher angular resolution using the {\it
  Herschel}/SPIRE \citep{Pilbratt:10}  on a subset of the {\it Planck} candidate
  DSFGs, consisting of the 228 brightest {\it Planck}
  sources. Details and initial results are discussed elsewhere
\citep{ Planck:15, Planck:16}.  Relevantly, 15 of the 228 sources are
discovered to be individual DSFGs boosted in brightness as a result of
gravitational lensing \citep{Canameras:15, Planck:15}.

\begin{figure}[t]
\includegraphics[width=\linewidth]{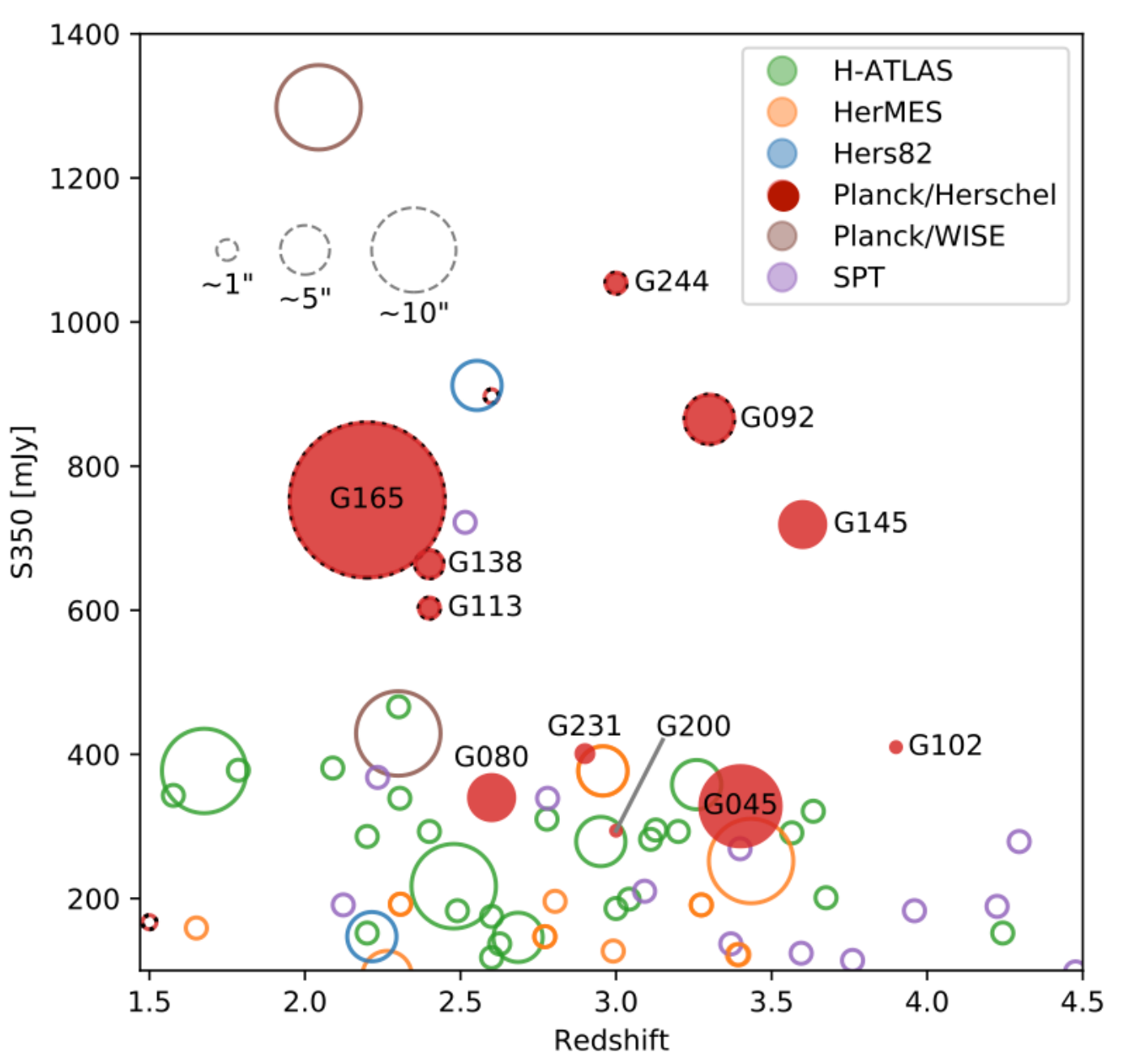}
\caption{$S_{350}$ flux density  of {\it lensed} DSFGs for {\bf the} {\it Planck/Herschel}-selected sample
\citep[filled red disks, ][]{Canameras:15}, complemented by others from the literature 
(open circles).  Our sample partially intersects with another survey of  {\it Planck/Herschel}-selected sources \citep[black-and-red dashed circles;][]{Harrington:16}.   The other lensed 
DSFGs indicated are sourced from the literature.  These are:  H-ATLAS \citep[green;][]{Harris:12, Bussmann:13, Calanog:14, 
Negrello:17}; HerMES \citep[orange;][]{Bussmann:13, Wardlow:13,  Calanog:14, 
Nayyeri:16}; Hers82 \citep[blue;][]{Nayyeri:16}; Planck/WISE \citep[brown;][]{Diaz-Sanchez:17}, and SPT \citep[purple;][]{Vieira:13, Weiss:13, Spilker:16}.  
In all cases, the symbol size represents the Einstein radius as estimated from the 
resolved image of the lensed source.  With some exceptions, the {\it Planck/Herschel} 
selection tends to  select lensed DSFGs with higher flux densities and larger 
Einstein radii.  
\label{fig1}}
\end{figure}

Herein we present new Hubble Space Telescope (HST) imaging and lensing analysis for 6 of the 15 
strongly lensed {\it Planck/Herschel}-selected sources.  We expand our study 
about one particular field in our sample, namely, PLCK G165.7+67.0 (hereafter 
G165), which shows strong-lensing constraints in the form of 
giant arcs and image multiplicities.
To better understand the properties of this one field, we acquire multi-wavelength 
imaging and spectroscopic follow-up observations which will be discussed in detail.

This paper is organized as follows.  In \S2 we compare our sample of strongly lensed 
DSFGs with others in the literature.  In \S3, we present new {\it HST} imaging data for our 
sample of six {\it Planck/Herschel}-selected strongly lensed DSFGs.  We also 
present new ground- and space-based observations of G165.  In \S4, we describe 
the data reduction and analysis of the follow-up data obtained for G165.  In \S5, we 
construct the strong-lensing model for G165.  This analysis is followed by a discussion 
of G165 in \S6,  in which we make independent determinations of  the mass, the
lensing strength, and the properties of the low inferred cluster gas pressure.  In \S7, 
we summarize our results.  Appendices are provided to describe the imaging and lensing 
analysis of all six fields in our \hst \ sample.  
We assume throughout a $\Lambda$CDM cosmology with 
$H_0=67$ km s$^{-1}$ Mpc$^{-1}$,  $\Omega_{m,0} = 0.32$,  and
$\Omega_{\Lambda,0} = 0.68$ {\bf \citep{PlanckVI:18}.}

\section{Strongly-Lensed DSFGs}
 
Although the details of the search strategies for strongly lensed DSFGs differ, 
most  algorithms set a high 350~$\mu$m flux density ($S_{350}$), or a high 
500~$\mu$m flux density ($S_{500}$), cut of 100 mJy.  To date, dozens 
of strongly lensed DSFGs in the redshift range 2\,$<$\,$z$\,$<$\,4 satisfy these 
criteria.  In Figure~1, we assemble the set of lensed DSFGs for the 
surveys, or subsets thereof, for which there are {\it Planck} $S_{350}$ 
flux densities,  spectroscopic redshifts for the lens and the lensed sources, 
reported Einstein radii, and images of the lensed sources.  For each DSFG 
in Figure~1, the symbol size is proportional to the size of the Einstein radius. 
For our sample, we measured the Einstein radius at the source redshift using 
our light-traces-mass model \citep{Zitrin:09b,Zitrin:15}.  For the objects
in other samples, we estimated the Einstein radius by eye from
the resolved image of the lensed source, or by a table made available to us for the case of the 
SPT sources (Spilker, private communication).  If a resolved image was not supplied
for a lensed source, then it is not included in Figure~1.  For reference, the Einstein radii are
assigned to either a 1$^{\prime \prime}$, 5$^{\prime \prime}$, or
10$^{\prime \prime}$ bin.  These bins are used to distinguish the scale of the lens type
approximately as a massive galaxy lens 
($M$\,$\sim$\,10$^{11}M_{\odot}$), a galaxy group lens ($M$\,$\sim$\,10$^{13}M_{\odot}$), or galaxy 
cluster lens ($M$\,$\sim$\,10$^{15}M_{\odot}$), respectively.  The legend gives the 
color-coded references and the bin sizes.  The brightest lensed DSFG, the ``Cosmic Eyebrow" 
($z = 2.0439$), stands out for its high submillimeter flux density 
($S_{350}$\,=\,1298\,$\pm$\,200 mJy).   
It was found by cross-correlating the sources in the WISE all-sky source catalog 
``AllWISE,"  with infrared-bright galaxies in the 
{\it Planck} compact source catalog 
\citep[single brown circle in Figure~1;][]{Diaz-Sanchez:17}.  Note that the submillimeter 
flux density of the Cosmic Eyebrow is measured from {\it Planck}/HFI data, which have
a higher uncertainty than the {\it Herschel}/SPIRE photometry 
used for the other comparison samples in Figure~1.  
Even if the true value is 
closer to its lower limit, it would still be the brightest known DSFG at
350 $\mu$m.

In the redshift range 2\,$<$\,$z$\,$<$\,4, a search for lensed DSFGs within the  
{\it Herschel} Astrophysical Terahertz Large Area Survey (H-ATLAS) using $S_{500}$
as a discriminator yields 22 lensed DSFGs covering a search area of 14.4 deg$^2$
\citep{Harris:12, Bussmann:13, Calanog:14, Negrello:17}.  This same approach applied
to the Herschel Multi-tiered Extragalactic Survey (HerMES) field extends the areal 
coverage by a factor of $\sim$7, resulting in 13 new lensed  sources
\citep{Bussmann:13, Wardlow:13, Calanog:14, Nayyeri:16}.  By applying similar flux 
density cuts to the Herschel Stripe 82 Survey (Hers82), an additional three lensed 
DSFGs are found \citep{Nayyeri:16}. 

From the ground, South Pole Telescope (SPT) data enable the selection of 
strongly lensed DSFGs based on the ratio of flux densities at 1.4 and 2.0 mm 
$S_{1.4mm}/S_{2mm}$,  which are consistent with thermal emission by dusty 
galaxies \citep{Vieira:10, Carlstrom:11}.  The brightest sources in the sample are 
then followed up at higher resolution using most notably the SubMillimeter Array 
(SMA) and the Atacama Large Millimeter Array (ALMA).  A total of 26 strongly lensed 
DSFGs are identified, which tend to be at higher redshifts  owing to their selection at 
longer wavelengths, and whose identifications are typically explained as galaxy--galaxy 
lensing events \citep{Weiss:13,Vieira:13}. 

From space, {\it Planck} and Herschel Space Observatory ({\it Herschel})
data are used to extend the search for lensed DSFGs to all available sky.  The search 
technique relies on the detection of the rest-frame far-infrared thermal dust peak,
which is a salient feature of DSFGs.  A strict lower limit on the flux density is imposed 
amounting to 600 mJy at 545 GHz to select only the most extreme sources. The 
expectation is that the brightest sources that are also compact at {\it Planck} resolution, 
and that remain compact in the {\it Herschel} follow-up observations are too faint to be 
explained by a single-field DSFG.  These sources are most likely  (1) multiple DSFGs, 
or (2) a single strongly lensed DSFG.  At the higher resolution using 
{\it Herschel}/SPIRE, the vast majority of sources resolve out into clumps of several 
submillimeter-bright objects in close projected proximity \citep{Planck:15, Planck:16}.  
These are the candidate galaxy overdense regions, which are potentially the 
high-redshift predecessors of massive lensing clusters at lower redshifts 
\citep{Planck:15, Flores-Cacho:16,Martinache:18, Kneissl:18}.   

At the same time, a small minority of 15 of 228 sources remained isolated, while 
also meeting additional flux density thresholds of $S_{350} > 300$ mJy and/or 
$S_{500} > 300$ mJy.  These sources show signatures of individual DSFGs that are 
boosted in brightness as a result of strong lensing.  Of these, 11 sources could be 
followed up at higher resolution using observing facilities from the Northern Hemisphere.  
Spectroscopic measurements of the lens and source redshifts, and identification of 
giant arc structures, strengthen  their lensing interpretation 
\citep[filled red disks in Figure~1,][]{Canameras:15, Nesvadba:16, Canameras:17a, Canameras:17b}.

The \citet{Harrington:16} sample (black-and-red dashed circles) is closely related 
to the \citet{Canameras:15} sample.  Their selection also relies on color using a 
combination of {\it Planck} and {\it Herschel}, yet the intersection is incomplete owing 
to the use of different {\it Planck} catalogs.  \citet{Harrington:16} select sources by cross-correlating 
{\it Herschel}/SPIRE with  {\it Planck} PCCS (six candidates), {\it Planck} 
HerMes (one candidate), and {\it Planck} HerS-82  (one candidate).   The selection of the 
{\it Planck/Herschel} sample
\citep{Canameras:15}  was made by applying color criteria to  {\it Planck} 
PCCS  and  {\it Herschel}/SPIRE (six candidates), and to {\it Planck} OT2 and  
{\it Herschel}/SPIRE (five candidates).  In all, three of eight of the \citet{Harrington:16} 
lensed DSFGs are new.

\begin{deluxetable}{cccc}
\tablecaption{The \hst \ Sample:  Observing Details}
\tablecolumns{3}
\tablehead{
\colhead{{\bf Lensing Field}}
& \colhead{Exp.\,(s)} & \colhead{Exp.\,(s)} \\
\colhead{}
& \colhead{F110W} & \colhead {F160W}
}
\startdata
PLCK G145.2+50.9 (G145) & 
2808 & 2736 \\
PLCK G244.8+54.9 (G244) & 
2592 &  2484 \\
PLCK G165.7+67.0 (G165) & 
2664 &  2556 \\
PLCK G045.1+61.1 (G045) & 
2556 &  2556 \\
PLCK G080.2+49.8 (G080) & 
2664 & 2664 \\
PLCK G092.5+42.9 (G092) & 
2808 & 2736 \\
\enddata
\end{deluxetable}

In conclusion, there is a tendency for {\it Planck/Herschel}-selected sources to have higher flux 
densities and larger Einstein radii than those drawn from the literature.   The cluster scale
of the lens may partially
explain this difference, in that a larger magnification factor ($\mu$) can be achieved, especially 
in the case of an Einstein ring such that $\mu \propto \sqrt{M_{lens}}$, where $M_{lens}$ is the 
mass of the lens.  The wider areal coverage of a factor of  $\sim$10 relative to the SPT and a 
factor of $\sim$100 or more relative to H-ATLAS, HerMES and Hers82 surveys also helps by 
allowing to set higher flux density thresholds{\bf ,} resulting in the identification of larger lenses in some 
cases.

\section{\sc Observations and Reduction}

We present new observations using {\it HST}/WFC3-IR for the six fields in our sample. 
 {\it HST}/WFC3-IR provides a high spatial resolution $0\farcs16$ FWHM at 
1.6 $\mu$m, and a high sensitivity with a reported 5$\sigma$ point source limiting 
magnitude in the F160W band of 27.0 AB mag \citep{Windhorst:11}.
We expand our study in the field of G165, selected because it produces
significantly more lensing
evidence, which leads to a more 
robust lens model.  To better characterize the additional lensing constraints in this 
one rich field, we acquired also new observations using LBT/LUCI\,+\,ARGOS, 
{\it Spitzer}/IRAC, {\it Gemini}/GMOS, and MMT/Hectospec.

\begin{figure*}
\centering
\includegraphics[width=1.0\linewidth]{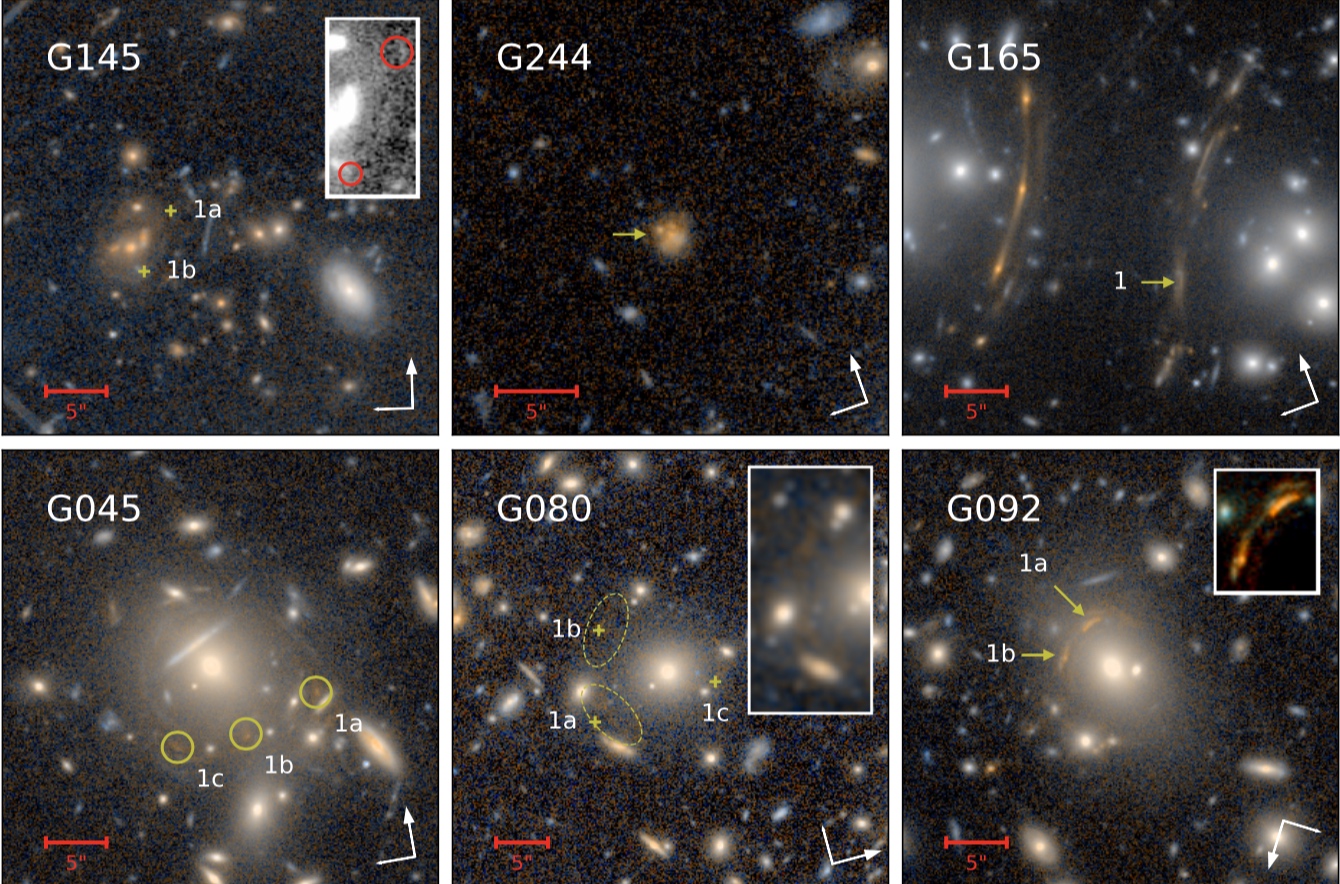}
\caption{
\hst \ F110W+F160W color composites of the central regions of the six fields 
in our  sample.  In all cases{\bf ,} we locate the NIR counterparts of the {\it Planck/Herschel} 
detections, where the gold plus signs mark the positions of the submillimeter 
sources from \citet{Canameras:15}.  The  images of the individual DSFGs appear 
in multiple locations in four of our fields:  G145, G165, G045, and G080.   For G092, 
the two NIR counterparts are not consistent with being counterimages, despite having 
similar colors, and are more likely to be separate, possibly interacting 
submillimeter-bright galaxies at a similar source redshift (labeled as `1a' and `1b'). 
For G244, the DSFG is a $1\farcs4$ partial Einstein ring that is resolved in the ALMA 
imaging \citep{Canameras:17a, Canameras:17b}.  This arc appears in the NIR as an 
extended source, which is blended with the main lensing galaxy such that giant arcs 
are not detected in these \hst \ data.  The cutout boxes in the upper right corners show 
the smoothed image (for G080), the single-band image (for G145), and the galaxy-halo-subtracted 
image (for G092), which more clearly shows the low surface brightness 
arcs.  In each panel, north is in the direction of the compass arrow and east is 
counterclockwise with respect to north.  A $5^{\prime\prime}$  scale bar 
is shown in the lower left corner of each panel for reference.
  \label{fig2}}
\end{figure*}

\begin{deluxetable*}{lcccccccc}
\tablecaption{The Sample of \hst \ Lensed DSFGs  
}
\tablehead{
\colhead{Arc ID}
&\colhead{R.\,A.}
&\colhead{Decl.}
&\colhead{F110W$_{AB}$}
&\colhead{F160W$_{AB}$}
&\colhead{Magnification}
&\colhead{Lens Size$^a$}
&\multicolumn2c{Spectroscopic Redshift}\\
\colhead{}
&\colhead{(J2000)}
&\colhead{(J2000)}
&\colhead{}
&\colhead{}
&\colhead{($\mu$)}
&\colhead{(arcsec)}
&\colhead{$z_{lens}$}
&\colhead{$z_{DSFG}$}
}
\startdata
G145\_DSFG\_1a  & 10:53:22.250 & +60:51:48.95 & $>$26.9 & $>$26.2  &12$\pm$1 &5.9 & 0.837$^b$ & 3.6$^c$\\
G145\_DSFG\_1b  & 10:53:22.563 & +60:51:44.03 &  $>$26.9 &$>$26.2   & 5$\pm$1 & $^{\prime\prime}$ & $^{\prime\prime}$ & $^{\prime\prime}$ \\
G244\_DSFG\_1   & 10:53:53.107 & +05:56:18.44 & $22.1^{+0.1}_{-0.1}$ & $21.0^{+0.1}_{-0.1}$ & 7-36$^d$ & 1.4$^d$& 1.525$^d$ & 3.005$^d$ \\
G165\_DSFG\_1a  & 11:27:14.731 & +42:28:22.56 & $23.0^{+0.1}_{-0.1}$ &  $22.2^{+0.1}_{-0.1}$ & $\apg 30$ & 13 & 0.351$^b$ & 2.2357$^e$ \\
G165\_DSFG\_1b  & 11:27:13.917 & +42:28:35.56 & $>$26.5  & $>25.6$ & $\sim$8 &$^{\prime\prime}$ &  $^{\prime\prime}$ & $^{\prime\prime}$ \\
G045\_DSFG\_1a  & 15:02:36.012 & +29:20:50.51 &  $>$26.9  & $25.5^{+0.2}_{-0.2}$ & $>$9$^f$& 8 & 0.549$^c$ & 3.427$^g$ \\
G045\_DSFG\_1b  & 15:02:36.479 & +29:20:47.90 &  $>$27.0  & $25.3^{+0.2}_{-0.2}$ & $>$9$^f$&$^{\prime\prime}$ &  $^{\prime\prime}$ &  $^{\prime\prime}$  \\
G045\_DSFG\_1c  & 15:02:36.921 & +29:20:47.96 & $>$26.7  & $25.8^{+0.3}_{-0.3}$ & $>$7$^f$ & $^{\prime\prime}$ &  $^{\prime\prime}$ &  $^{\prime\prime}$\\
G080\_DSFG\_1a   & 15:44:33.202 & +50:23:43.53 & $>$27.1 & $>$26.5  &$\sim$20  &7 & 0.670$^c$ & 2.6$^c$ \\
G080\_DSFG\_1b   & 15:44:32.483 & +50:23:41.69 & $>$27.5 & $>$26.5 &$\sim$20  & $^{\prime\prime}$ &  $^{\prime\prime}$ &  $^{\prime\prime}$\\
G092\_DSFG\_1a$^g$  & 16:09:17.842 & +60:45:19.41 & $24.2^{+0.1}_{-0.1}$ & $23.0^{+0.1}_{-0.1}$ & $\sim$20& {\bf --} & 0.45$^i$ & 3.256$^e$\\
G092\_DSFG\_1b$^g$  & 16:09:17.693 & +60:45:22.31 & $25.0^{+0.2}_{-0.2}$ & $23.7^{+0.1}_{-0.1}$ & $\sim$20&{\bf --}  &  $^{\prime\prime}$ &  $^{\prime\prime}$\\
\enddata
\tablenotetext{a}{The effective Einstein radius is reported, defined as 
$\theta_E = \sqrt{A/\pi}$, where 
$A$ is the area inside the 
critical curve \citep[e.g.,][]{Acebron:18}.}
\tablenotetext{b}{The spectroscopic redshift comes from this paper.}
\tablenotetext{c}{The spectroscopic redshift comes from \citet{Canameras:15}.}
\tablenotetext{d}{The lensed image is only partially resolved in our \hst \ image (Figure 2).  These values for the magnification factor, the size of the Einstein radius, and the redshift are drawn from 
the ALMA data in \citet{Canameras:17a, Canameras:17b}.}
\tablenotetext{e}{The spectroscopic redshift comes from \citet{Harrington:16}.}
\tablenotetext{f}{Magnification estimates for the three brightest emission-line regions 
of this arclet family were first reported in \citet{Nesvadba:16} as $\mu$ = 10\,-\,22.}
\tablenotetext{g}{The spectroscopic redshift comes from \citet{Nesvadba:16}.}
\tablenotetext{h}{ We find that these two images are likely to originate from two different sources at a similar redshift.}
\tablenotetext{i}{The spectroscopic redshift comes from SDSS DR14 data.}
\end{deluxetable*}

\subsection{\sc \hst \  Observations}

We obtained imaging of six {\it Planck/Herschel}-selected 
fields between 2015 December and 2016 July with
the \hst \ Wide Field Camera 3 IR detector (WFC3-IR) in Cycle 23 (GO-14233; PI:  Frye).  The fields are: PLCK G145.2+50.9 (G145), PLCK G244.8+54.9 (G244), PLCK G165.7+67.0 (G165), PLCK G045.1+61.1 (G045), PLCK G080.2+49.8 (G080), and PLCK G092.5+42.9 (G092).
The imaging is composed of one orbit each in the F110W and F160W filters.
Table~1 gives the observing details.

The  WFC3-IR images are redrizzled using the software package \textit{DrizzlePac} 
(Fruchter et al.~2010).   We adopt values for the photon-sensitive effective size of a 
pixel to its real size ({\tt final\_pixfrac}), and a final pixel scale ({\tt final\_scale}), of 
$0.85$ and {\bf $0\farcs06$}, respectively.  We checked the redrizzled images by 
computing the FWHM of a few stars in each field.  In some cases, such as the G165 
field, there were only two isolated, unsaturated stars within the field of view, so we
substituted a compact and isolated elliptical galaxy as a third source.  Redrizzling of
the data in each case resulted in a 3\,--\,10\% improvement in image quality (FWHM) over the pipeline products.
The final reduced images reach comparable depths to the CLASH clusters.  For one representative case, 
G165, we compute 10$\sigma$ limiting magnitudes of
 F110W$_{AB}$\,=\,26.9 mag and F160W$_{AB}$\,=\,26.2 mag for point sources inside 0$\farcs$4 apertures. 
We find that the image depth and filters
are sufficient to make NIR detections of the 
strongly lensed DSFG in each of our sample fields. We also identify other examples of giant arcs and/or image multiplicities
in some cases.  In Figure 2, we 
present the \hst \ color images of the central regions for each of the six fields.
We refer the reader to Appendix A for further details regarding the search for the 
NIR
counterparts.

The WFC3 F160W images are used as detection images for the matched aperture photometry.  We custom-built our code to cope with the unusual morphologies
peculiar to arcs in the central regions of massive lensing clusters.  
We detect sources by applying a $\sigma$-clipping algorithm with respect to the local background RMS values. The local background, in turn, is estimated following a similar approach to that in \textit{SExtractor} \citep{Bertin:96}.  Briefly, the image is divided into patches of $100\times 100$ pixels, with 
the background in each patch represented by the $\sigma$-clipped median.  The local background is then estimated by a smooth $3\times 3$ spline interpolation over these patches. To ensure robust detection of objects, we smooth the image with a Gaussian filter of FWHM\,$\sim0\farcs$2.
Objects are deblended using the watershed PYTHON algorithm in {\tt astropy.photutils}
\citep{Astropy:18}.  Artifacts such as diffraction spikes are visually identified and removed.

We assign apertures to each galaxy image by measuring
the semi-major/minor axis sizes at $3$ times the FWHM lengths of the detected objects, typically
amounting to 0$\farcs$6. Elliptical annuli are used to get the best estimate of the background, with an inner radius equal to the photometric aperture and an outer radius equal
to $1.2$ times the inner radius. We compute the aperture flux and 
then subtract the area-scaled local background level 
within the annuli. The flux uncertainty is computed as the quadratic sum of the local smooth background RMS value only.  We note that aperture corrections are minimal owing to our large extraction aperture
of $>$\,$0\farcs4$.
In three fields, G145, G045, and G080, the multiple images of the single DSFG are 
typically too faint
and/or blended with the halos of 
bright cluster members near in projection to measure a flux.  In these cases, a 1$\sigma$ upper limit on their fluxes
is reported.  In Table~2 we present the
photometric catalog of the lensed DSFGs for our sample.  The magnification factor,
$\mu$, of the lensed DSFG was measured using our light-traces-mass lens model.  The 
effective Einstein radius at the redshift of the lensed source, $\theta_E = \sqrt{A/\pi}$,
was measured from our lens model.

\subsection{\sc LBT Observations}

We acquired imaging of G165 in $K$ band using the LBT Advanced Rayleigh 
Ground layer adaptive Optics System \citep[ARGOS;][]{Rabien:19}  during instrument commissioning 
time in December 2016 (2016B, PI:  Frye).   ARGOS corrects ground-layer 
distortions in the imaging of the two 8.4 m apertures using two three-beam constellation 
lasers as guide stars that are fixed 
to each aperture.   The ARGOS instrumentation operates through the LUCI 
 imager and multislit spectrometer.  High-quality corrections of up to FWHM 
$\approx$\,0$\farcs$25 at $K$ are achievable across a large field of view 
($4^{\prime} \times 4^{\prime}$)
at a native pixel scale of 
0$\farcs$12\,pix$^{-1}$.  We acquired  LBT/LUCI\,+\,ARGOS data in monocular 
mode on two separate nights: 46 minutes of observation 
using LUCI2 on 9 December and 42 minutes  using LUCI1 on 2016 December 15.  
We note that the LUCI1 set of observations have a slightly 
shorter exposure time and, in turn, a slightly higher per-pixel RMS uncertainty. However, they 
yield a fainter point-source detection limit owing to the lower FWHM as measured in  
isolated and unsaturated stars.
We choose to analyze
the data separately from each night and only to
combine the photometric measurements at the last step.

  \begin{figure}[h!]
\includegraphics[width=\linewidth]{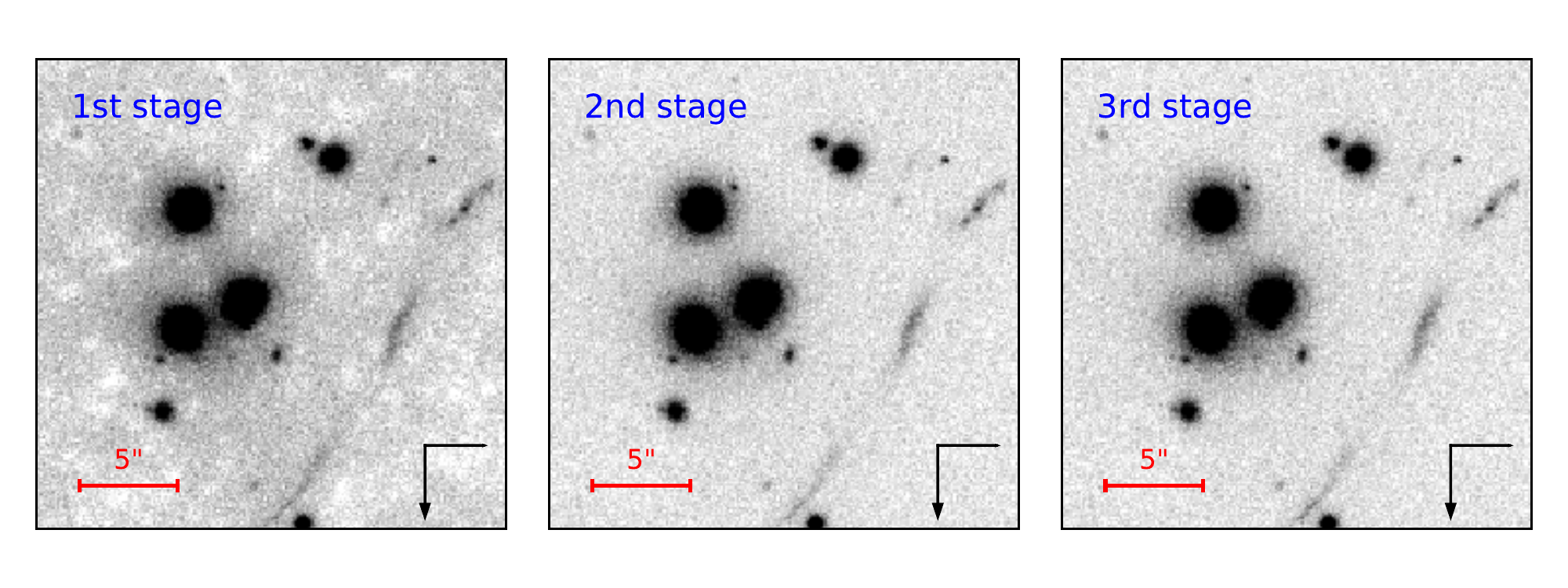}
\caption{Background subtraction of our LBT/LUCI\,+\,ARGOS $K$ band imaging data is accomplished
in an iterative process as illustrated above for a small region near the cluster center.
The difference in image flatness between the stacked images after the first stage,  and the 
final image upon extending the bright object masks 
(second and third stages), is evident.  We sample the background by placing test boxes down 
that are isolated from bright sources.  We find the flatness to improve, or alternatively for the rms level of the 
background to decrease, by typically 9\% between the first and second stages 
and to converge after the third stage.  
In each panel north is down and east is to the right. 
 \label{fig3}}
\end{figure}

Random dithers of up to 40$^{\prime \prime}$ are imposed between individual
exposures to optimize the sky subtraction 
in the crowded cluster regions and to eliminate detector artifacts.  Such large dithers 
require high point-source stability across the field.  As a check, we estimate the pointing 
error at each dither position by stacking the object frames, and then measuring the 
positional centroids of 13 pre-selected cluster members that span the full field of view,
and that are isolated from bright sources.  We find that the typical translational shift 
between images is $\sim$\,2 pixels, or $\sim$\,0$\farcs$25, with negligible rotation.  The WCS 
information in each object frame is updated accordingly.  At this point we resample the 
images onto the same pixel grid using the flux-conserved, overlapping pixel-area method 
in the Python routine {\tt astropy.reprojection}, as needed.

We design our own reduction pipeline to ensure high flatness
across the chip and to maximize the signal-to-noise ratio of the data.  As a first step, we 
subtract the dark frames from all object frames.  We then proceed to find
the  best estimate of the background.  
Within a single exposure, the sky background varies by $\sim$100 ADUs, 
comparable to the integrated flux of some of the fainter cluster members.  Therefore, instead of creating sky frames by taking the median at multiple pointings directly, we 
apply a ``normalizing-rescaling'' approach, in which we construct the master sky frames 
 using exposures adjacent in time,  which are 
then scaled to the background level of each object frame prior to the subtraction.

We designate each dither pointing ``$i$" as the co-addition of 24 individual object frames in 
$5$ s exposures, all taken at the same position (120 s total science time){\bf ,} plus $\sim$\,0~s 
readout time owing to nondestructive readouts.  The 5~s exposure was chosen to be small to avoid persistence and nonlinearity effects.  
We find that a reasonable compromise in image combination is to
collect the temporally closest five dither pointings about each {\it i}th dither pointing in a running 
boxcar,  equating to a total clock time of $\sim$14 minutes including overheads.  The result of including more dither pointings is a slight improvement in 
the background noise but a degradation in image flatness. We mask out the bright sources in all the frames
of the running boxcar to avoid biasing 
the result{\bf ,} or ``master-background{\bf ,}" upward of its true value with unwanted cluster halo light.  Before dividing this master frame into the {\it i}th dither pointing,
we divide the  dither pointing frame
by its five-pass, 3$\sigma$ clipped median value\footnote{When computing the $\sigma$-clipped statistics of an image, 
we cut the borders by 300 pixels to minimize the bias due to the hot/cold edges of 
infrared detectors.} to obtain the mean image of these ``normalized''
 object frames. 
As a last step, we scale this new running master background frame to the 3-$\sigma$ clipped 
median of this $i$th dither  pointing to match the sky-background level at 
the exact time of the exposure.  Our background-subtracted dither pointings yield the
``first-stage" result shown in Figure~3.

\begin{figure}[h!]
\centering
\includegraphics[width=\linewidth]{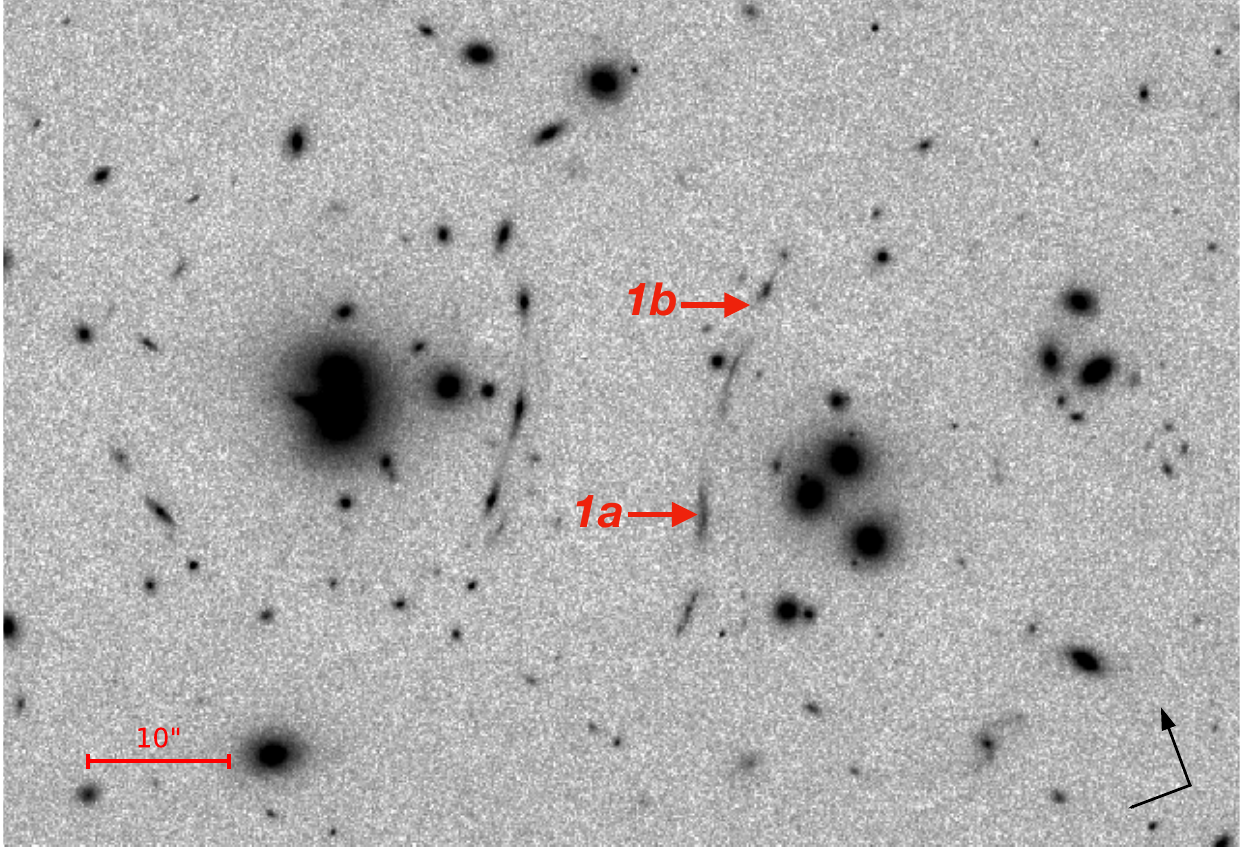}
\caption{High-resolution $K$-band image (FWHM\,$\approx$\,$0\farcs29$) for the central region  of G165, using LBT/LUCI\,+\,ARGOS.  Dozens of strongly lensed galaxy images and arclet families
are detected.  The lensed galaxy DSFG\_1a is spatially resolved (labeled as ``1a").  
Notably, 
we identify a new image that we designate as its
counterimage,  {\it DSFG\_1b} (labeled as ``1b").    See Figure~5 for a 3-band color image which includes these data 
alongside our \hst \ data set.  North is in the direction of the compass arrow, and east is to the left.    
A $10^{\prime\prime}$  scale bar is shown in the lower left corner for reference. 
 \label{fig4}}
\end{figure}

Following an iterative approach, we introduce two additional stages to the background subtraction, 
each time using the previous stage result as a starting point.  The main difference is that we continue 
to extend the bright object mask into the fainter outskirts of the masked sources.  We avoid aggressively 
expanding the bright source mask, as increasing the number of masked pixels improves the flatness,
but 
at the expense of  the noise level, as fewer frames are available from which to estimate the 
background.  The ``third-stage" result is obtained by performing another iteration of the ``second-stage" result.
To assess image quality after each reduction stage, we compute the 
background RMS values inside of seven test boxes of size $50\,\times\,200$ pixels 
located in regions isolated from bright sources.  We find the background RMS to 
decrease on average by 9 \% following the first stage and to converge to the 
$\apll$1\% level following the second stage (see Figure~3). 

 To make corrections for pixel-to-pixel variations, 
we first tried applying a flat field to the data in the usual way  prior to subtracting off the background.
The result was unsatisfactory because image artifacts remained in the data.  On reversing 
the order of these two operations, we found an 
improvement in the image flatness and the removal of the image artifacts. This improvement arises
because our master flat field is constructed by combining {\it local} sky frames generated
as a natural part of our background subtraction algorithm.  Finally, we stacked the sky-subtracted and flat-fielded object frames to produce our final data product.  We report a mean 
$K$-band FWHM of 0$\farcs$53  for the 2016 December 9 run (LUCI2), and 0$\farcs$29  
for the 2016 December 15 run (LUCI1).   In all, for the 2016 December 9 and the 2016
December 15 runs,
we reach a 10$\sigma$ limiting magnitude of  $K_{AB}$\,=\,22.6 mag and $K_{AB}$\,=\,23.5 mag
inside apertures of 4\,$\times$\,FWHM, respectively.  We do not combine the final images 
from the two different detectors, because the 2016 December 15 data have better spatial resolution
(Figure~4).  We emphasize, however, that our photometry is measured using both nights of data
and weighted
by an inverse-variance-weighted mean of the two fluxes, as described in \S4.1.

The large aperture of LBT and the high resolution of LUCI+ARGOS 
have enabled rare detections of dozens of arcs in a ground-based image.  
Moreover,
a side-by-side comparison of our $K$-band image and our \hst \ WFC-IR images allows identifications of single galaxy
images that appear in multiple positions in the image plane, or ``arclet families."  
By  this approach{\bf ,} we confirm the image position of the lensed DSFG, {\it Arc 1a}, and
make the first detection of its counterimage, {\it Arc~1b}, at its model-predicted
location (Figure~5).  Until now, we were unable to detect {\it Arc~1b} in our  {\it HST} images,
as the arc was too faint and {\bf too} red.  The $K$-band 
detection of this doubly imaged source with a known spectroscopic redshift 
is impactful because it allows us to break the
mass-sheet degeneracy to construct a robust mass map.   
In all, 11 arclet family
members are detected in five separate arclet families.  
Thus, the LBT/LUCI+ARGOS $K$ imaging data effectively extend the 
wavelength reach of {\it HST}, thereby opening up new discovery space that favors the redder galaxy
populations.   For additional details regarding the performance of the LBT/LUCI+ARGOS instrument,
see \citet{Rabien:19}.

\subsection{\sc Spitzer Observations}

We acquired imaging in the G165 field on 2016 February 2 using the {\it Spitzer} InfraRed
Camera (IRAC) in the  3.6 $\mu$m and 4.5 $\mu$m channels as part of a larger
program  (Cy13, GO-13024; PI:  Yan) to image the fields of massive lensing clusters that would make good targets
for JWST.
The on-target exposure time was set to 12,000 s in each of the two channels. The Spitzer Science Center (SSC) 
processed these data using the standard SSC
pipeline, and we made the final image mosaics based on these
products. A detailed analysis of the data set of this entire program will appear
in a separate paper (Yan et al.~2018, in preparation).  We also refer the reader to
the description in \citet{Griffiths:18}, where the reduction of their IRAC data from
the same {\it Spitzer} program is discussed. In Figure~5, we
show the 3.6\,$\mu$m IRAC mosaic of this field (right panel).  The
two images of the DSFG, {\it G165\_DSFG\_1a} and {\it G165\_DSFG\_1b}, are both bright,
with $S_{3.6,AB}$\,$\approx$\,19.1 mag and $S_{3.6,AB}$\,$\sim$\,19.9 mag, respectively.

\subsection{\sc MMT Observations}

We obtained spectroscopy in the field of G165 on 2015 February 14 using  MMT/Hectospec  \citep{Fabricant:13},
as a part of a larger program (2015A; PI:  Frye). 
Hectospec is a multifiber spectrograph that  assigns optical fibers on
the sky with minimum allowed separations of $\apg$20$\farcs$
To maximize the wavelength range for measuring redshifts, we selected the 270 groove mm$^{-1}$ grism, which 
covers a wavelength range of 3700\,--\,9150~\AA\  at a dispersion of 1.21 \AA \ pixel$^{-1}$.  
We chose to position 23 fibers (20 galaxy targets plus 3 standard stars) with priorities set to the positions of the brighter examples of prominent giant arcs and cluster members,  as selected by their NIR photometric redshift estimates 
made using Canada France Hawaii Telescope (CFHT) plus {\it Spitzer}/IRAC imaging.   We
refer to
\citet{Canameras:18} for details on the photometric analysis and photometric redshifts.  
Note that as the planning for this observing run took place prior to receiving the \hst \ data set,
we were not able to fine-tune the target list to include new arclet family members.
The observations were composed of a single Hectospec pointing with 7\,$\times$\,1020~s exposures taken under variable seeing conditions of $\sim$\,1\,--\,2$\farcs$   This was sufficient for our science goal given the 1$\farcs$5 fiber widths and 
{\bf the} relatively bright 
magnitudes of the targets of $i_{AB}\simeq 18$\,--\,22 mag.

The reductions proceeded in a standard fashion
using the IDL/Hectospec Reduction Software package (HSRED) obtained from the Smithsonian Astrophysical Observatory Telescope Data Center.\footnote{(https://www.mmto.org/node/536)}
We removed cosmic rays using the code ``LA Cosmic" \citep{vanDokkum:01}.   
Corrections for pixel-to-pixel variations, fringe corrections, and fiber identifications
are accomplished using a dome flat.
Background subtractions were made after first averaging together the spectra set to blank-sky positions,
taken under the same conditions as the science data.  
The wavelength solution is found in two ways, using both an {HeNeAr} lamp exposure and
the positions of prominent night-sky lines.
 We co-added the individual exposures to yield the final  reduced spectra.

\begin{figure}[h!]
\includegraphics[width=\linewidth]{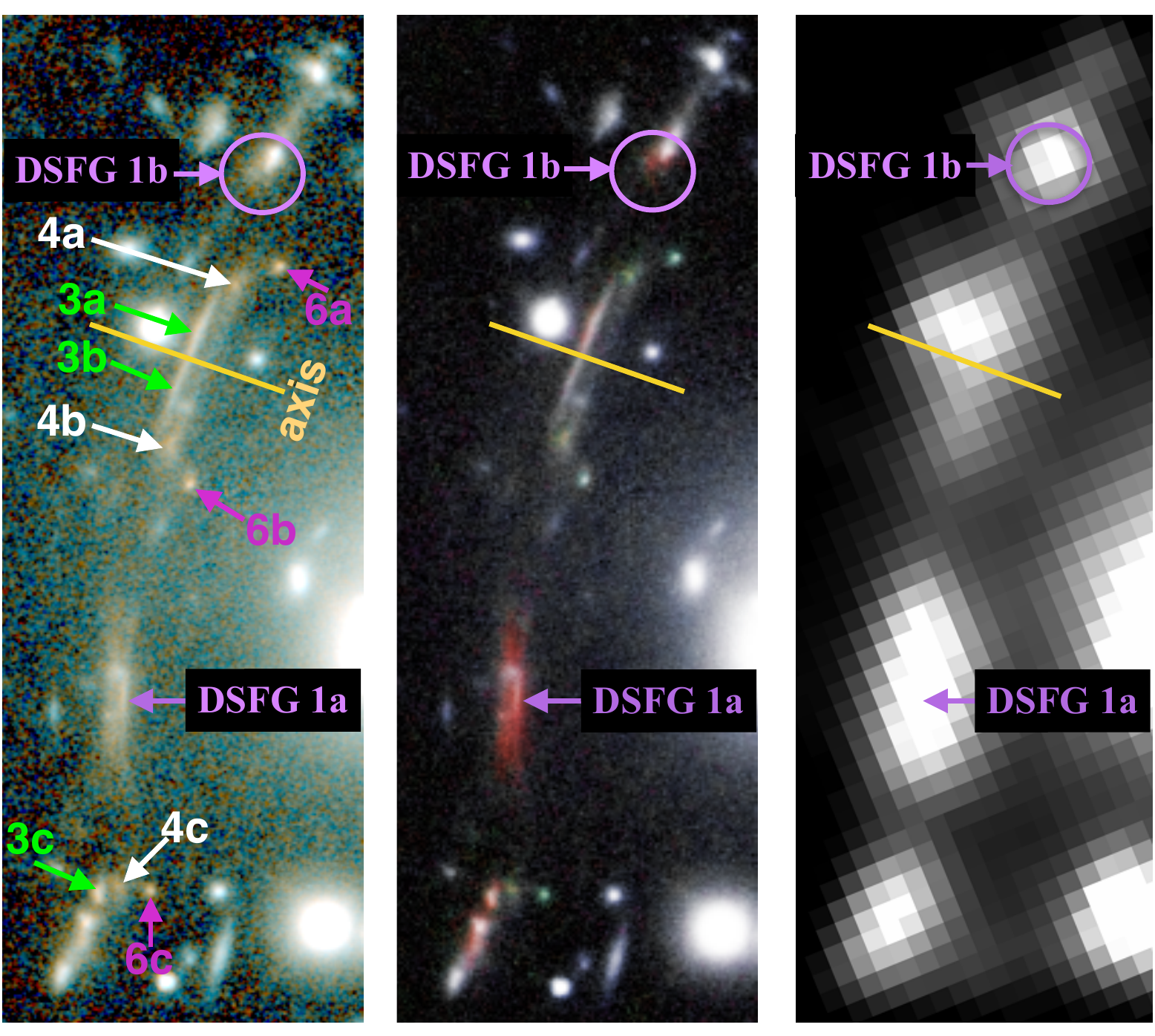}
\caption{ 
Left:   two-band color image of a central region of the G165 field in \hst \ WFC3-IR
F110W+F160W.  DSFG\_1a appears at the expected location based on the 
submillimeter imaging \citep{Canameras:15}.  Sharing this crowded 
central region are three other arclet families, which all show fold images 
about an axis of symmetry, as labeled.  Middle:   three-band image of the
 \hst \ WFC3-IR F110W (blue), F160W (green) and LBT/LUCI\,+\,ARGOS $K$ (red) data.
DSFG\_1a stands out on account of its red color and is punctuated by two blue compact images, which may potentially be two images of compact star forming regions
arising from a single object in the background (see \S~6.2).  Our lens model also predicts for 
there to be a fainter counterimage, DSFG\_1b, which is discovered in the $K$-band
data at its model-predicted location.
Right:{\it Spitzer}/IRAC image in the 3.6 $\mu$m band.  DSFG\_1a and DSFG\_1b 
are both bright, with $S_{3.6,AB}$\,$\approx$\,19.1 mag and $S_{3.6,AB}$\,$\sim$\,19.9 mag, 
respectively.   All images are 10$^{\prime \prime}$\,$\times$\,30$^{\prime \prime}$ on a side and have the 
same orientation as in Figure~\ref{fig2}.
}
\end{figure}

Secure spectroscopic measurements are made for 19 objects, which we define as
the high-significance detection 
of two or more spectral features  ($>$2$\sigma$ level in the continuum).  Our catalog results in measurements for five
 new cluster members with $z$\,=\,0.326--0.376, and 11 new sources with redshifts  0.388\,$<$\,$z$\,$<$\,0.622.  Three galaxies have MMT/Hectospec redshifts that place them in the foreground.  
See \S 4.2 for additional details and the redshift catalog.

\begin{deluxetable*}{cllcccc}
\tablecaption{The PLCK\_G165.7+67.0
(G165) Lensed DSFG and the Arclet Families}
\tablehead{
 \colhead{Arc}
&\colhead{R.\,A.}
&\colhead{Decl.}
&\colhead{F110W$_{AB}$}
&\colhead{F160W$_{AB}$}
&\colhead{$K_{AB}$} 
&\colhead{{\bf ${z_{pred}}^a$}}\\
 \colhead{ID}
&\colhead{(J2000)}
&\colhead{(J2000)}
&\colhead{(mag)}
&\colhead{(mag)}
&\colhead{(mag)}
&\colhead{}
}
\startdata
G165\_DSFG\_1a&11:27:14.731&+42:28:22.56&$23.0^{+0.2}_{-0.2}$&$22.2^{+0.2}_{-0.2}$&$18.9^{+0.2}_{-0.2}$&\hspace{0.1cm}2.2357$^b$\\
G165\_DSFG\_1b&11:27:13.917&+42:28:35.54&$>26.5$                    &$>25.6$                    &$22.6^{+0.2}_{-0.2}$&2.2357  \\
G165\_2a & 11:27:15.962  & +42:28:29.00  & $22.8^{+0.1}_{-0.1}$ & $21.4^{+0.1}_{-0.1}$ & $18.5^{+0.1}_{-0.1}$  &1.7 \\
G165\_2b & 11:27:15.606  & +42:28:34.18  & $22.8^{+0.1}_{-0.1}$ & $21.3^{+0.1}_{-0.1}$ & $18.3^{+0.1}_{-0.1}$  &$^{\prime \prime}$\\
G165\_2c & 11:27:15.325  & +42:28:41.32  & $22.8^{+0.1}_{-0.1}$ & $21.4^{+0.1}_{-0.1}$ & $18.5^{+0.1}_{-0.1}$  &$^{\prime \prime}$ \\
G165\_3a & 11:27:14.146  & +42:28:32.00  & $25.0^{+0.1}_{-0.1}$ & $24.3^{+0.1}_{-0.1}$ & $21.6^{+0.1}_{-0.1}$  &2.7 \\
G165\_3b & 11:27:14.330  & +42:28:30.36  & $25.6^{+0.2}_{-0.2}$ & $25.1^{+0.2}_{-0.2}$ & $22.3^{+0.3}_{-0.2}$  &$^{\prime \prime}$  \\
G165\_3c & 11:27:14.969  & +42:28:17.34  & $25.3^{+0.2}_{-0.2}$ & $24.6^{+0.2}_{-0.2}$ & $21.9^{+0.3}_{-0.3}$  &$^{\prime \prime}$ \\
G165\_4a & 11:27:14.059  & +42:28:32.73  & $26.1^{+0.3}_{-0.2}$ & $24.8^{+0.1}_{-0.1}$ & $22.7^{+0.4}_{-0.3}$  &2.7\\
G165\_4b & 11:27:14.372  & +42:28:29.00  & $25.8^{+0.2}_{-0.2}$ & $24.7^{+0.2}_{-0.2}$ & $22.2^{+0.2}_{-0.2}$  &$^{\prime \prime}$ \\
G165\_4c & 11:27:14.909  & +42:28:17.33  & $>$26.9                     & $26.0^{+0.2}_{-0.2}$ & $>$23.6                      &$^{\prime \prime}$ \\
G165\_5a & 11:27:13.187  & +42:28:25.83  & $25.8^{+0.2}_{-0.2}$ & $25.7^{+0.3}_{-0.2}$ & $>$23.0                      &4.3\\
G165\_5b & 11:27:13.188  & +42:28:24.67  & $25.7^{+0.2}_{-0.2}$ & $25.5^{+0.2}_{-0.2}$ & $>$23.4                      &$^{\prime \prime}$ \\
G165\_6a & 11:27:13.924  & +42:28:32.79  & $25.8^{+0.2}_{-0.2}$ & $24.7^{+0.1}_{-0.1}$ & $>$23.2                    &2.7\\
G165\_6b & 11:27:14.358  & +42:28:27.72  & $25.8^{+0.2}_{-0.2}$ & $24.5^{+0.1}_{-0.1}$ & $23.2^{+0.7}_{-0.4}$&$^{\prime \prime}$  \\
G165\_6c & 11:27:14.836  & +42:28:16.88  & $>$26.3                     & $25.1^{+0.2}_{-0.2}$ & $>$23.2                    &$^{\prime \prime}$  \\
G165\_7a & 11:27:15.300  & +42:28:38.35  & $>$26.2                     & $25.6^{+0.4}_{-0.3}$ & $>$23.5                    &1.7 \\
G165\_7b & 11:27:15.397  & +42:28:35.96  & $>$26.9                     & $25.2^{+0.3}_{-0.2}$ & $>$23.5                    &$^{\prime \prime}$  \\
G165\_7c & 11:27:16.083  & +42:28:25.84  & $26.2^{+0.4}_{-0.3}$  & $25.7^{+0.5}_{-0.4}$ & $>$23.5                    &$^{\prime \prime}$ \\
G165\_8a & 11:27:15.210  & +42:28:40.85  & $>$26.0                      & $>$25.3                    & $>$23.0                    &1.7 \\
G165\_8b & 11:27:15.581  & +42:28:32.37  & $>$26.3                      & $>$25.5                    & $>$23.0                    &$^{\prime \prime}$ \\
G165\_8c & 11:27:15.834  & +42:28:28.36  & $>$26.8                       & $>$26.2                   & $>$23.0                    &$^{\prime \prime}$ \\
G165\_9a & 11:27:15.423  & +42:28:40.59  & $26.5^{+0.4}_{-0.3}$   & $>$25.6                   & $>$22.9                    &1.5 \\
G165\_9b & 11:27:15.568  & +42:28:36.06  & $>$26.4                       & $>$25.6                   & $>$22.4                    &$^{\prime \prime}$ \\
G165\_9c & 11:27:16.096  & +42:28:28.03  & $>$26.6                       & $>$25.6                   & $>$22.7                     &$^{\prime \prime}$ \\
G165\_10a & 11:27:15.171  & +42:28:38.92 & $>$26.3                      & $26.2^{+0.5}_{-0.3}$& $>$22.8                    &1.7 \\
G165\_10b & 11:27:15.453  & +42:28:32.90 & $26.5^{+0.3}_{-0.3}$  & $25.9^{+0.3}_{-0.3}$ & $>$23.0                   &$^{\prime \prime}$\\
G165\_11a & 11:27:15.761  & +42:28:42.28 & $24.7^{+0.1}_{-0.1}$  & $24.5^{+0.1}_{-0.1}$ & $>$22.0                   &2.1\\
G165\_11b & 11:27:15.784  & +42:28:40.45 & $25.0^{+0.1}_{-0.1}$  & $24.8^{+0.2}_{-0.2}$ & $>$22.0                   &$^{\prime \prime}$ \\
G165\_11c & 11:27:16.510  & +42:28:26.12 & $25.0^{+0.1}_{-0.1}$   & $24.7^{+0.1}_{-0.1}$ &$>$22.7                   &$^{\prime \prime}$ \\
\enddata
\tablenotetext{a}{Unless otherwise noted, these redshifts are predictions drawn from our strong-lensing model that await spectroscopic confirmation.}
\tablenotetext{b}{The spectroscopic redshift for G165\_DSFG\_1a comes from \citet{Harrington:16}.}
\label{t_G165fams}
\end{deluxetable*}

\subsection{\sc Gemini Observations}

We obtained further spectroscopy in the field of G165
using the {\it Gemini}-North Multi-Object spectrograph (GMOS) as a part of a larger program (GN-2016A-Q-30; PI:  Frye).   
The observations took place on 2016 April 27.
We selected the B600 line mm$^{-1}$ grating, which has a wavelength coverage measured
from our data of a total of 2975 \AA \ about the central wavelength for each slitlet
 at a dispersion of 0.92 \AA \ pixel$^{-1}$.  As we  
did not have the \hst \ images  in time to plan this observing run, 
we populated the slit masks first with prominent arcs selected from the CFHT
image from \citet{Canameras:15}, followed by 
cluster members selected from our {\it Gemini} pre-imaging data.  We chose
 1$^{\prime \prime}$ slits to match typical
seeing on-site.  We acquired six science exposures of 1200~s, two each at central wavelengths of 645~nm, 650~nm, and 655~nm to correct for chip gaps.  Arc spectra were obtained within $\pm$\,1 night of the observations using the CuAr lamps at similar central wavelengths.  Dispersed flat-fields were taken at each of the three central wavelength (and hence grating tilt) configurations.

The initial calibrations of bias subtraction and flat-fielding proceeded in the standard way 
using the IRAF {\it Gemini} reduction package.\footnote{IRAF is distributed by the National 
Optical Astronomy Observatories, which are operated by the Association of Universities for 
Research in Astronomy, Inc., under cooperative agreement with the National Science 
Foundation.}  We removed cosmic rays prior to the background subtraction using the IRAF 
task GEMCRSPEC.  For the wavelength calibration, there is a tendency for the IRAF algorithm 
to introduce wavelength offsets of the stacked spatial rows, especially for the smaller spectral 
``boxes."   To avoid introducing this undesirable spatial feature into the data,  we chose 
instead to use a pipeline written in IDL by one of us (B.\,L.\,F.).  The IDL pipeline includes the tasks 
mentioned below and is discussed elsewhere \citep{Frye:02, Frye:07, Frye:08}.  Briefly, the IDL 
pipeline avoids repixelization by identifying the flexure-induced instrumental curvature imprinted 
onto the individual spectrum box edges between the 2D spectra.  This curvature amounts to 1-3 pixel shifts from the
center to edge of the CCD, which are easily fit by low-order polynomials.  We then 
wavelength-calibrate the data in two ways:  using the arc lamps and using the night-sky lines.  
As both outputs had an RMS on the wavelength fit of $<$\,0.5~\AA, we choose to use the sky lines 
for the potential benefit  that the wavelength  references are embedded directly onto the data at 
the time of the observations.

Cosmic-ray hits on the object were removed in 1D by a comparison 
of the stacked spectra from the six different exposures using our IDL 
task SPADD \citep{Frye:02, Frye:07, Frye:08}.  Thresholds are set for 
the acceptable number of cosmic-ray hits per pixel in the stack to avoid 
removal of real spectral features.   We measured redshifts for the 1D 
co-added spectra using our IDL task SPEC   \citep{Frye:02, Frye:07, Frye:08}.
Our catalog results in spectroscopic measurements for 32 galaxies in the G165 field.  Of these,
we  find nine cluster members that are new with 0.326\,$<$\,$z$\,$<$\,0.376 and 18 new lensed sources with 0.386\,$<$\,$z$\,$<$\,1.065.
Five galaxies have {\it Gemini}/GMOS redshifts that place them in the foreground of the lens.  
We refer to \S 4.2 for additional details and the full redshift catalog.


\section{\sc Analysis and Results for G165}

We describe our algorithms for performing the matched photometry for the \hst \ 
plus LBT imaging. We then analyze the combined results of the MMT,
{\it Gemini}, and archival ground-based spectroscopy.

 \subsection{\sc The Photometry}

To include the LBT/LUCI\,+\,ARGOS data in our catalog alongside the \hst \ photometry, we first translate the central locations of our photometric apertures defined by the F160W image onto the $K$ band using the WCS information.
Although the FWHM resolution of our LUCI1 $K$-band data ($\mathrm{FWHM}\sim\,0\farcs$29) is higher than that of our two \hst \ bands (0$\farcs$22 and 0$\farcs$18 for F110W and F160W
bands, respectively),
we do not alter the aperture sizes and ellipticities, as there is adequate matching
 to detect the vast majority of the sources.
The data from LUCI1 and LUCI2 are obtained under different weather conditions, and the field orientation angles and plate scales are slightly different for LUCI1 and LUCI2.  As a result, we opt to conduct $K$-band photometry separately for LUCI1 and LUCI2 images and only 
afterward to compute the aperture fluxes by applying an inverse-variance-weighted mean of the two values.

Table~3 gives the complete photometric catalog for all 11  arclet families.  
As the photometric depth at $K$ is shallower than for the \hst \ $J$- and $H$-band
data, the aperture fluxes for some sources and arclets fall below their 1-$\sigma$ 
uncertainties.  In such cases, we report the detection limit of the aperture 
fluxes. The redshift, $z_{pred}$, gives the predicted value for the redshift using our lens 
model.  Notably, 11 arclet family members are detected in our LBT LUCI+ARGOS 
$K$-band image.  They are {\it Arcs 1a, 1b}, {\it Arcs 2a, 2b, 2c}, {\it Arcs 3a, 3b, 3c}, 
{\it Arcs 4a, 4b, 4c}, and {\it Arc 6b}.
The lensed DSFG, {\it G165\_DSFG\_1a},  has 
F110W$_{AB}$ = 23.0$^{+0.2}_{-0.2}$ mag, F160W$_{AB}$ = 22.2$^{+0.2}_{-0.2}$ mag, 
and K$_{AB}$ = 18.9$^{+0.2}_{-0.2}$ mag, bright enough to make ground-based 
spectroscopic follow-up feasible.

\begin{figure}[h!]
\includegraphics[width=\linewidth]{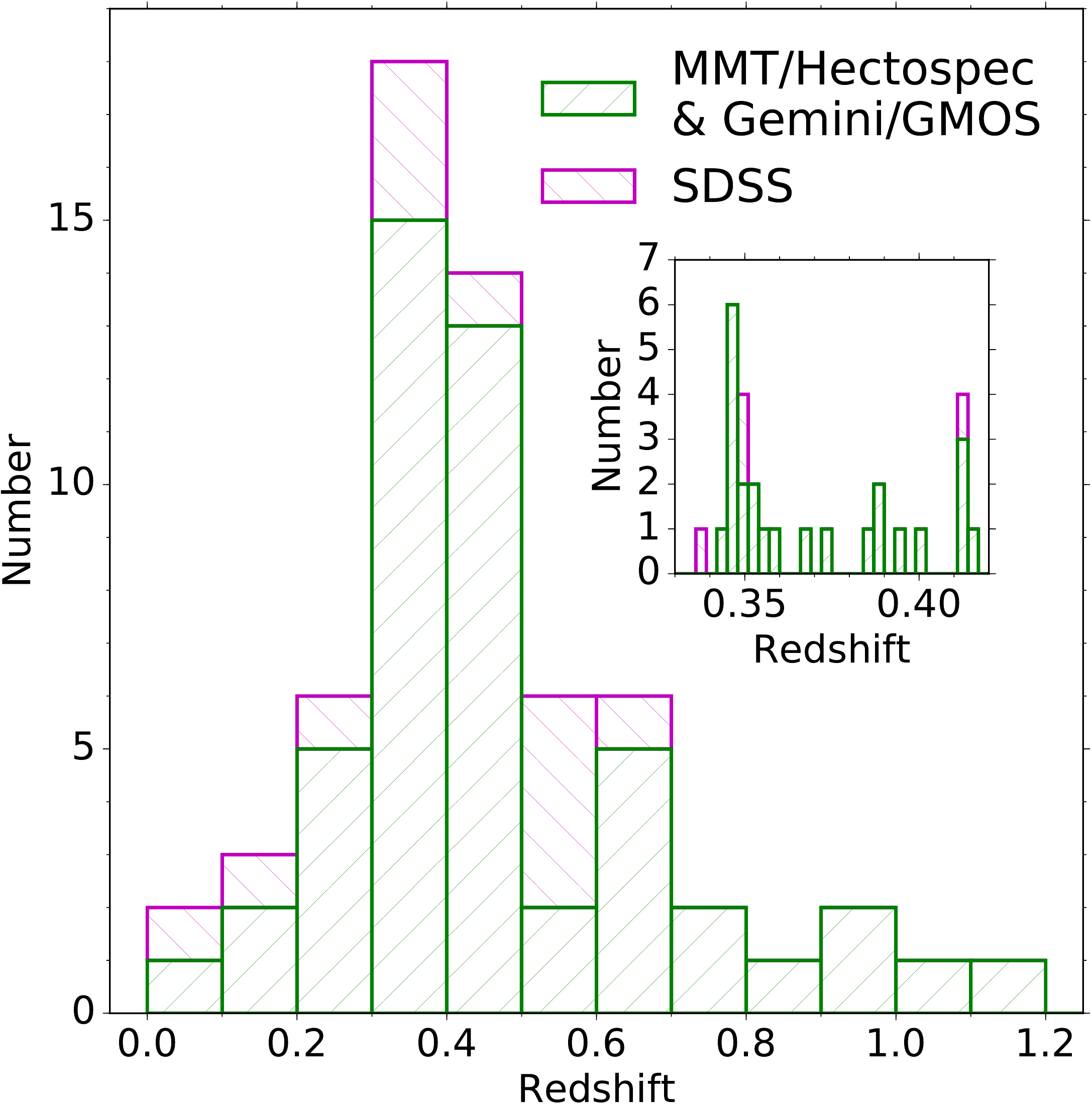}
\caption{Histogram of spectroscopic redshifts in the G165 field.  The redshift catalog
combines  results from our MMT/Hectospec and {\it Gemini}/GMOS data sets (green), and 
objects drawn from the literature (SDSS, DR 14; magenta).  We measure a value for the lens 
redshift of $z$\,=\,0.351, based on 18 cluster members in the range of 
0.326\,$<$\,$z$\,$<$\,0.376.  The inset histogram peaks at the lens redshift and shows 
a secondary peak at higher redshift  that may indicate the presence of  an unrelated
background galaxy group. These cluster members range in position from the center out 
to a cluster-centric radius of $\approx0.8$ Mpc.}
\end{figure}

\startlongtable
\begin{deluxetable*}{lccccccc}
\tablecaption{{\bf Spectroscopic} Redshifts in the G165 Field$^a$}
\tablecolumns{6}
\tablewidth{0pc}
\tablehead{
  \colhead{Source ID}
& \colhead{R.\,A.}
& \colhead{Decl.}
& \colhead{{\bf $z_{spec}$}}
& \colhead{Ref}
& \colhead{$i_{\mathrm{SDSS,AB}}$}
& \colhead{F110W$_{AB}$}
& \colhead{F160W$_{AB}$}\\
  \colhead{}
& \colhead{(J2000)}
& \colhead{(J2000)}
& \colhead{}
& \colhead{}
& \colhead{(mag)}
& \colhead{(mag)}
& \colhead{(mag)}
}
\startdata
s1      & 11:26:36.173  &  +42:30:08.42 & 0.621 & S & 18.99 & --    & -- \\
s2      & 11:26:43.968  &  +42:31:05.16 & 0.274 & S & 18.17 & --    & -- \\
s3      & 11:26:45.732  &  +42:28:15.65 & 0.121 & S & 17.05 & --    & -- \\
s4      & 11:26:46.387  &  +42:26:51.81 & 0.412 & S & 18.76 & --    & -- \\
s5      & 11:26:48.850  &  +42:28:33.05 & 0.471 & S & 19.59 & --    & -- \\
s6      & 11:26:59.151  &  +42:30:11.10 & 0.412 & H & 21.34 & --    & -- \\
s7      & 11:26:59.471  &  +42:28:10.81 & 0.346 & H & 18.74 & --    & -- \\ 
s8      & 11:26:59.999  &  +42:27:03.71 & 0.388 & H & 19.77 & --    & -- \\
s9      & 11:27:00.233  &  +42:31:03.06 & 0.232 & H & 17.84 & --    & -- \\
s10     & 11:27:02.557  &  +42:29:12.35 & 0.445 & H & 19.68 & --    & -- \\
s11     & 11:27:02.667  &  +42:27:20.19 & 0.348 & G & 20.02 & --    & -- \\ 
s12     & 11:27:04.239 &  +42:29:32.38 & 0.399 & G & 21.46 & --    & -- \\
s13     & 11:27:04.785 &  +42:31:21.53 & 0.389 & H & 20.15 & --    & -- \\
s14     & 11:27:05.558  &  +42:25:55.06 & 0.445 & G & 21.42 & --    & -- \\
s15     & 11:27:05.754 &  +42:27:34.60 & 0.623 & G & --    & --    & -- \\
s16     & 11:27:06.732  &  +42:27:50.40 & 0.275 & G & 17.50 & --    & -- \\ 
s17     & 11:27:06.787 &  +42:27:23.25 & 0.623 & G & —-    & --    & -- \\
s18     & 11:27:07.552 &  +42:28:22.50 & 0.622 & H & 20.35 & --    & -- \\
s19     & 11:27:09.420 &  +42:30:38.23 & 0.624 & G & 21.01 & --    & -- \\
s20     & 11:27:09.564  &  +42:30:10.90 & 0.358 & G & --   & --    & -- \\ 
s21     & 11:27:10.774  &  +42:30:14.15 & 0.033 & S & 14.71 & --    & -- \\
s22     & 11:27:11.137 &  +42:26:50.88 & 0.412 & H & 18.56 & --    & -- \\
s23     & 11:27:12.283 &  +42:28:23.88 & 0.353 & H & 21.08 & 20.2 & 19.9  \\ 
s24     & 11:27:13.046 &  +42:27:09.58 & 0.386 & G & 21.25 & 21.4 & 21.3 \\ 
s25     & 11:27:13.133 &  +42:31:09.47 & 0.510 & S & 19.29 & --    & -- \\
s26     & 11:27:13.300 &  +42:30:27.68 & 0.347 & G & 21.30 & --    & -- \\ 
s27     & 11:27:13.444 &  +42:27:00.54 & 0.411 & G & 20.46 & --    & -- \\
s28     & 11:27:13.653 &  +42:30:39.21 & 0.374 & G & 21.29 & --    & -- \\ 
s29     & 11:27:13.680  &  +42:28:22.44 & 0.348 & S & 18.33 & 18.7 & 18.3 \\ 
s30     & 11:27:14.803 &  +42:27:37.58 & 0.135 & H & 19.06 & 18.6   & 18.2   \\ 
s31     & 11:27:15.312 &  +42:29:00.99 & 0.305 & G & 20.18 & 22.0 & 21.7  \\
s32     & 11:27:15.370 &  +42:27:35.60 & 1.065 & G & 19.06 & 22.0 & 21.9  \\
s33     & 11:27:16.596 &  +42:28:40.99 & 0.348 & S & 18.00 & 18.3 & 17.9 \\ 
s34     & 11:27:16.664 &  +42:27:23.07 & 0.720 & G & 22.24 & --    & -- \\
s35     & 11:27:16.692  &  +42:28:38.15 & 0.338 & S & 17.16 & --    & -- \\ 
s36     & 11:27:16.894 &  +42:31:08.83 & 0.508 & G & 21.35 & --    & -- \\
s37     & 11:27:17.145 &  +42:26:07.18 & 0.146 & G & 21.62 & --    & -- \\
s38     & 11:27:17.928  &  +42:27:20.43 & 0.193 & H & 18.41 & --    & -- \\
s39     & 11:27:18.027 &  +42:26:48.30 & 0.368 & G & 20.23 & --    & -- \\ 
s40     & 11:27:18.501  &  +42:26:02.94 & 0.623 & G & -€"-    & --    & -- \\
s41     & 11:27:18.594 &  +42:29:29.25 & 0.471 & G & --    & 22.9 & 22.7  \\
s42     & 11:27:18.652 &  +42:28:09.81 & 0.354 & G & --    & 21.6 & 21.3  \\ 
s43     & 11:27:18.879  &  +42:29:55.38 & 0.254 & G & 21.44 & --    & -- \\
s44     & 11:27:19.394 &  +42:29:50.95 & 0.346 & H & 20.28 & --    & -- \\ 
s45     & 11:27:19.452  &  +42:27:01.73 & 0.723 & G & -    & --    & -- \\
s46     & 11:27:19.908  &  +42:30:18.18 & 0.351 & G & 20.59 & --    & -- \\ 
s47     & 11:27:20.146  &  +42:29:18.46 & 0.275 & G & 20.40 & 24.2 & 24.0  \\
s48     & 11:27:20.379  &  +42:30:28.11 & 0.999 & G & 20.97 & --    & -- \\
s49     & 11:27:20.386  &  +42:30:51.64 & 0.443 & H & -€"-    & --    & -- \\
s50     & 11:27:20.458  &  +42:27:59.16 & 0.345 & H & 19.17 & 18.7 & 18.4 \\ 
s51     & 11:27:20.509 &  +42:29:01.78 & 0.414 & G & -    & --    & -- \\
s52     & 11:27:22.652  &  +42:31:08.80 & 0.344 & H & 19.54 & --    & -- \\ 
s53     & 11:27:23.369 &  +42:29:53.01 & 0.348 & G & 21.04 & --    & -- \\ 
s54     & 11:27:23.383 &  +42:26:27.56 & 0.914 & G & 20.12 & --    & -- \\
s55     & 11:27:23.685  &  +42:26:49.52 & 0.916 & G & --    & --    & -- \\
s56     & 11:27:23.833  &+42:28:42.64 & 0.0$^b$& H &...    & 18.5 & -- \\
s57     & 11:27:24.564  &  +42:29:48.99 & 0.347 & G & 21.06 & --    & -- \\ 
s58     & 11:27:24.695  &  +42:29:04.92 & 0.759 & G & 21.72 & --    & -- \\
s59     & 11:27:25.340  &  +42:27:43.93 & 0.395 & H & 22.12 & --    & -- \\
s60     & 11:27:26.521 &  +42:26:58.02 & 0.347 & H & 19.55 & --    & -- \\ 
s61     & 11:27:29.150  &  +42:30:23.45 & 0.544 & S & 19.36 & --    & -- \\
s62     & 11:27:31.872  &  +42:27:41.04 & 0.522 & S & 19.60 & --    & -- \\
\enddata
\tablenotetext{a}{This object is a star.}
\end{deluxetable*}

\subsection{{\bf {\sc The Spectroscopy: G165 Cluster Members}}}

The catalog for all 62 objects in the G165 field with measured redshifts 
is given in Table~4, and presented as a redshift histogram in Figure~6.  
All objects in our redshift catalog 
are secure, by which we mean that we require that two or more spectroscopic features be detected at the 
$\apg$2$\sigma$ level relative to the continuum.  In the case of a single emission line, we 
require also the detection of a second significant feature such as a continuum break.  
We did not encounter any lone emission lines in this census.  
Hypothetically, if we did detect a single emission line blueward of H$\alpha$ without a continuum break, 
then we would flag it as a Ly$\alpha$ candidate line.  This is because a single emission line shortward 
of the rest-frame wavelength of H$\alpha$ will likely be [\ion{O}{3}]~
$\lambda \lambda$4959,5007, [\ion{O}{2}]~$\lambda \lambda$3727, 3729, or Ly$\alpha$.  
In the first case, we would resolve both lines of the doublet.  In the second case, the redshift would 
be sufficiently small, $z$\,$\apll$\,0.8, that we would also detect [\ion{O}{3}]~$\lambda \lambda$4959, 5007 in our spectral 
bandpass for both the MMT/Hectospec and {\it Gemini}/GMOS data sets.  This would leave the likely 
identification of such a feature as Ly$\alpha$.  However, we do not detect any high-redshift candidates in this particular data set.

\begin{figure}[h!]
\includegraphics[width=\linewidth]{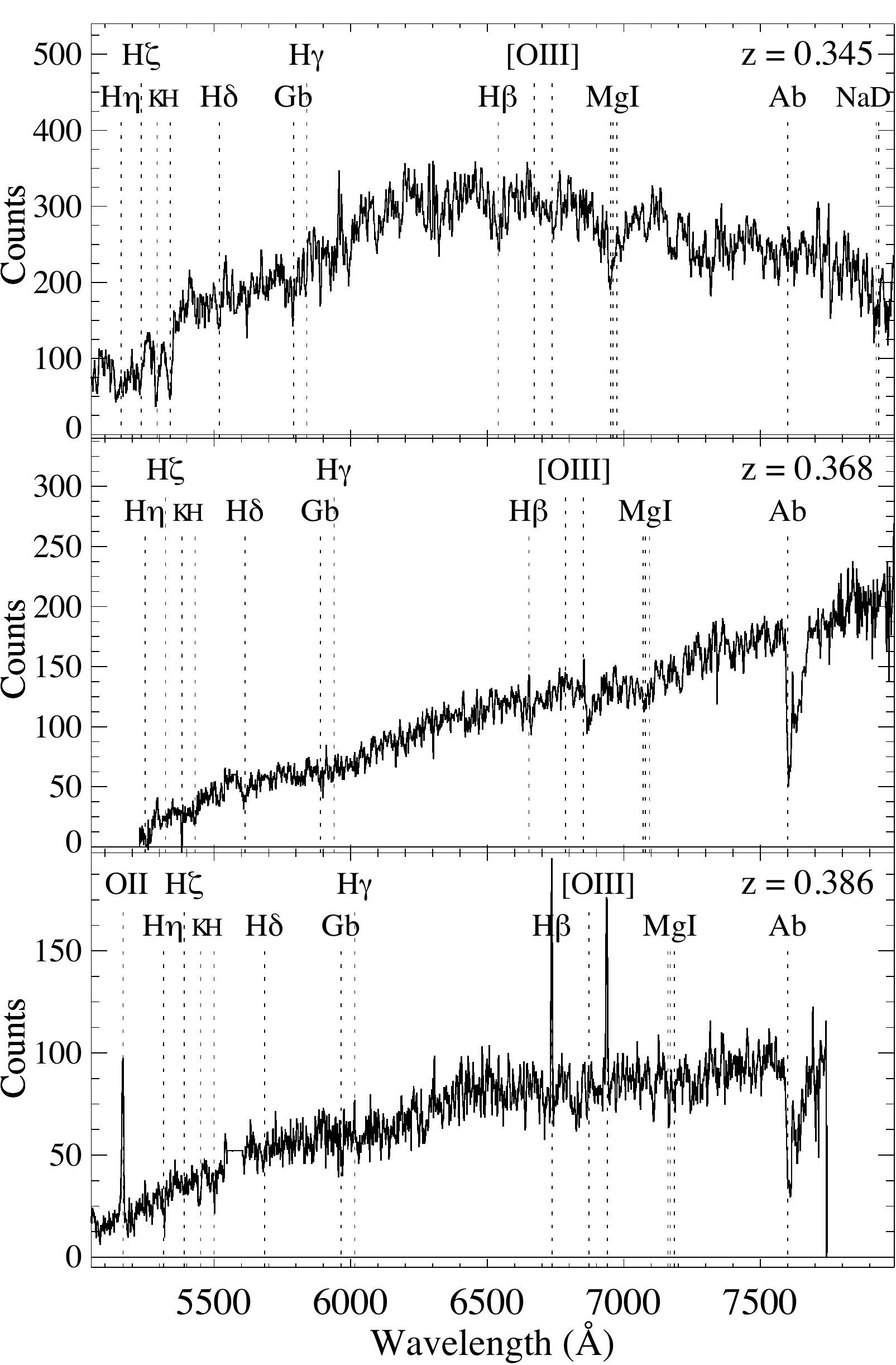}
\caption{
Sample spectra of cluster members in the field of G165.  Only 
spectroscopy that yields secure redshifts are used in this study.
The galaxies typically show the absorption features expected of
 early-type galaxies such as the MMT/Hectospec spectrum of Source ID ``s50" (top panel).  
 Other objects show
 a combination
of stellar absorption plus emission line features, such as the {\it Gemini}/GMOS spectrum of Source ID ``s39" 
(middle panel).   Three cluster members show  nebular emission line features, such as
the {\it Gemini}/GMOS spectrum of 
Source ID ``s24" (bottom panel).  See Table~4 for details.}
\end{figure}

Typical absorption- and emission-line features 
identified in our data (depending on the redshift and type) are:
\ion{Fe}{2}~$\lambda \lambda$2587, 2600, 
\ion{Mg}{2}~$\lambda \lambda$2796, 2803, \ion{Mg}{1}~$\lambda$2852, 
[\ion{O}{2}]~$\lambda \lambda$3727, 3729,  Ca~H \& K, G-band,  
[\ion{O}{3}]~$\lambda \lambda$4959, 5007, 
\ion{Mg}{1}~$\lambda \lambda \lambda$5167, 5173, 5184, NaD,  Balmer family (H$\alpha$ through H$\theta$), 
[\ion{S}{2}]~$\lambda \lambda$6716, 6731.  See Figure~7 for sample spectra.
Note that all 50 secure redshifts obtained from our spectroscopy and reported in this paper with an 
``H" for Hectospec or ``G" for Gemini/GMOS are new to the literature.

We specify cluster membership in the standard way, by requiring the  redshifts to be in the range 
0.326\,$<$\,$z$\,$<$\,0.376, which corresponds to $\pm$3$\sigma$ with respect to the mean of $z=0.351$.
In total, we have spectroscopic redshifts for 18 galaxies in the cluster.   
Of these,
six cluster members are drawn from our MMT/Hectospec data,
and an additional nine cluster members come from our {\it Gemini}/GMOS data.   The information
for the
remaining three cluster members comes from all other available sources, which for this field
is only SDSS (DR13).
Those cluster members in common with the smaller {\it HST} field fall reasonably well onto 
the red sequence of the color-magnitude diagram (CMD; see \S B.1 in the Appendix).   
Five galaxies with spectroscopic redshifts extending behind the cluster and in the range 
(0.412\,$\leq$\,$z$\,$\leq$\,0.414)
are not included
in this cluster member set, yet they may be members 
 of an unrelated background galaxy group  (Figure~6, inset).

To cope with the small sample size, we choose to measure the velocity dispersion using the
Gapper method (see \citet{Hou:09} for details). We compute a velocity dispersion of 
$\sigma$\,=\,2400\,$\pm$\,620~km~s$^{-1}$ from the 18 cluster members, corresponding to a large value 
for the virial mass of $M_V = (9.1 \pm 0.4)\,\times\,10^{15}$ $M_{\odot}$ within 1 Mpc.  If we now restrict 
the angular extent to match the scale of our \hst \ observations of  $\theta$ = 50$\farcc$ or 
$\approx$\,250\,kpc, then  13  cluster members are removed. 
The velocity dispersion for this smaller redshift set is $\sigma$ = 2000\,$\pm\,300$\,km s$^{-1}$, yielding 
again a large value for the mass of $M_V = 1.3 {^{+0.04}_{-0.70}} \times 10^{15}$ $M_{\odot}$.  The 
uncertainties on the velocity dispersion are found by summing up the uncertainties in the galaxy redshifts 
in quadrature.  
The value for the dynamical mass within the cluster core is not uncommon for massive clusters
\citep{Girardi:93},  and at the same time is higher than the mean value for  CLASH clusters by a factor 
of three \citep{Siegel:16}.  

We emphasize that our values for $\sigma$, and hence also for $M_V$,
will be biased upward relative to the true value if the 
line-of-sight velocities are enhanced relative to those in the transverse direction.  
It is relevant here to consider a nonspherical 
velocity structure, as a bimodal mass distribution is evident in the imaging data. 
The cluster galaxies separate out naturally into  
two main mass concentrations:  a northeast (NE), and a southwest (SW) region.
We take the cluster center to be situated at the center of this bimodal mass distribution,
with a positional uncertainty that depends on the relative masses.   Given that
each of the two mass regions produce similar numbers of arcs and arclet families, conservatively we expect the
mass ratio to be $\apll10$.  The uncertainty on the cluster center translates
into an uncertainty in the virial radius, and none of the above 
takes into account the potentially large systemic errors due to the unknown {\bf true} radial and
 velocity structure
of the cluster.
We return to the discussion of the cluster kinematics in \S~6.1.

\begin{figure*}[t!]
\includegraphics[width=\linewidth]{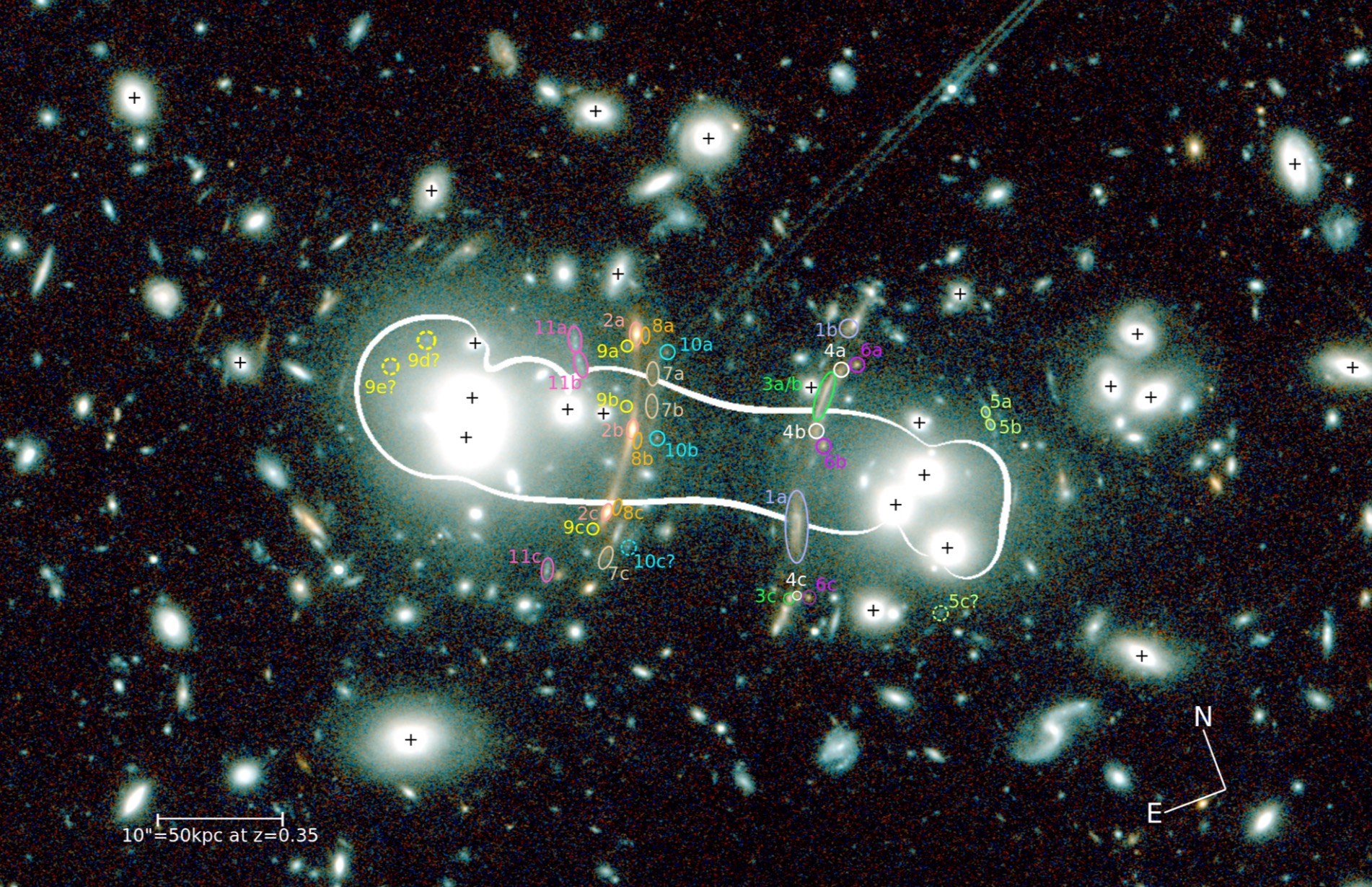}
\caption{{\it HST} color image of the G165 field.
We identify 11 new arclet families (labeled and color-coded) in this rich field.
The doubly imaged DSFG notably bisects the critical curve (lavender-colored ellipse ``1a"),
and appears again just north of the gravitational potential as ``1b."  The critical curve 
is obtained from our LTM model (see \S5), using the pipeline by \citep{Zitrin:09b, Zitrin:15}.
We make arclet family designations by a combination of their similar colors, 
morphologies,  image symmetries, and model-predicted locations (see Figure~9 
for image stamps).
 }
\end{figure*}

\section{\sc Strong-lens Modeling}

\subsection{The Approach}

We perform a strong-lensing analysis for the fields in our sample by an approach
that relies on the assumption that the light approximately traces the mass, or ``LTM," 
such that the galaxies are biased tracers of the dark matter.  A similar LTM 
methodology has been used to constrain the 2D mass distribution for cluster lenses 
extending back to some of the first examples of image multiplicities in cluster 
environments such as A2390 \citep{Frye:98}, and Cl0024 \citep{Broadhurst:00}.  This lensing analysis was 
subsequently extended to accommodate the properties of the first cluster field to 
show large numbers of arclet families, namely, the \hst \ Advanced Camera for 
Surveys (ACS) image of A1689 \citep{Broadhurst:05}.  To construct our mass maps, 
we use the well-tested implementation of the LTM pipeline by \citet{Zitrin:09b, Zitrin:15}. 
We also refer to \citet{Acebron:18} and \citet{Cibirka:18} for 
additional descriptions.  

In the LTM model, the lensing galaxies are assigned a power-law
mass density distribution scaled in proportion to their  luminosities.   The power-law 
index is left as a free parameter and is the same for all lensing galaxies.
The superposition of the mass distributions of the individual lensing galaxies, which  
makes up the initial 2D mass distribution, is then smoothed by a Gaussian kernel
to approximate the dark matter distribution, whose width is the second free parameter of the model.
The dominant dark matter and galaxy distributions are, in turn, 
summed up with a relative weight, which adds another a free parameter of the model, and then they are normalized (to a specific source redshift),
 which necessitates the fourth free parameter. 
Finally, the model accommodates a two-parameter external shear to provide 
additional flexibility.  The values for these six parameters are constrained
by the positions, orientations, and relative brightnesses of the arclet families.
The best-fit model and 
errors are optimized through a Monte-Carlo Markov Chain using thousands of steps.

 \begin{figure*}
\includegraphics[width=\linewidth]{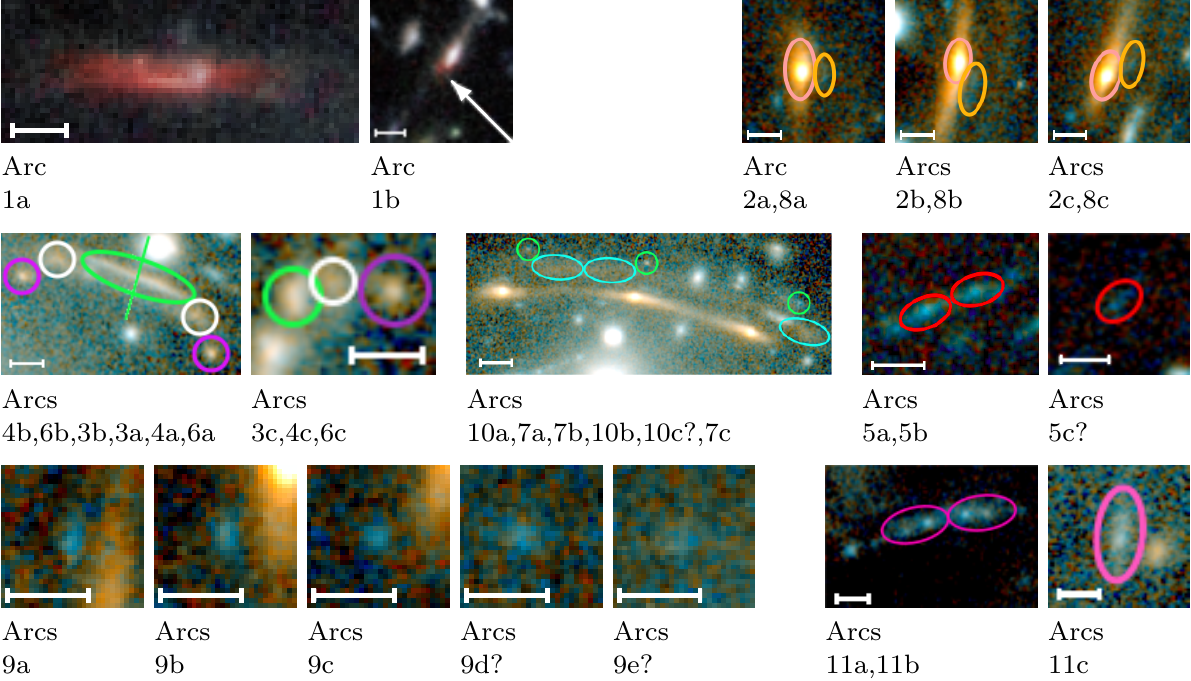}
\caption{Image stamps depicting each of the 11 arclet families in the G165 field, as labeled.  
Model predictions for the designation of extended arclet family members are also shown, 
qualified by a question mark. Except for {\it Arcs 1a} and {\it 1b}, the two \hst \ bands are shown, 
which are demonstrated here to be valuable in the identification of the arclet families by their 
similar colors.  For {\it Arcs 1a and 1b} we present the three-color image consisting of the HST 
WFC3-IR F110W and F160W bands and the LBT LUCI+ARGOS $K$-band.  For 
{\it Arcs 9d?} and {\it 9e?}, we first subtracted off the cluster light using {\it Galfit} prior to producing
the two-band color image.  The high-resolution $K$-band data set extends the reach of {\it HST} at comparable spatial resolution, thereby enabling 
the discovery of the ultrared counterimage of the lensed DSFG, {\it Arc 1b}, which
is also confirmed by our {\it Spitzer} image at 3.6 $\mu$m (Figure~5).  
The image stamps all have a 1$^{\prime \prime}$\,=\,5 kpc bar (assuming 10$^{\prime \prime}$\,=\,50 kpc) provided for reference.
The images maintain the same orientation as in Figure~2. }
\end{figure*}

The exquisite spatial resolution of \hst \ makes feasible the designation of arclet 
families based on morphology and color.  Moreover, the \hst \ images show 
obvious axes of symmetry superimposed onto the field (see example in Figure~5), which allow for 
the identification of image multiplicities even without the aid of measured redshifts 
in some cases.  
Arclet 
families, in turn, constrain the model by imposing the condition that each family 
member image originates from the same source.  The best-fit model is the one that 
minimizes the angular separations between the observed and predicted (relensed) image 
positions in the image plane.  Notably, in addition to providing confirmation of the 
locations of the counterimages, the strong-lensing model also has predictive 
power to locate new image counterparts that can be searched for in the data to
iteratively improve on the model result. 
Because secure spectroscopic redshifts are not available for every arclet family, the ratios of the relative angular diameter
distances of the lens to the source, $d_{LS}$, and of the observer to the source, $d_S$, are left as free parameters to be optimized in the minimization of the model. In such cases, we allow a wide range of 
relative values of $\frac{d_{LS}}{d_S}$\,=\,$\pm$\, 0.12, which equates roughly to a redshift range of 
$z$\,$\simeq$\,1.5\,--\,7.    We expect the redshifts of the arcs roughly to coincide with
the star-formation rate density peak of the universe of $z = 2$\,--\,3 \citep{Madau:14}.  Even so,
this broader redshift accounts for the potential outliers, as recommended in 
\citet{Johnson:16} for the case of limited spectroscopic redshift information and also found
to be a useful constraint in \citet{Cibirka:18}, and \citet{Cerny:18}.   

The lens model for the G165 cluster field  is discussed in  detail below, and a lensing analysis for the other five fields appears in 
 \S~B.1 of
the Appendix.  We emphasize that all arclet families discovered in this study are supported by our
LTM model.

\subsection{G165}

 The cluster members form a pattern on the CMD referred to as the red sequence \citep{Gladders:00}.  The G165 field shows a prominent 
red sequence characterized by very little scatter, making the designation of the 
cluster members especially straightforward by applying a color cut in the usual way.  
At the same time, the amount of scatter 
increases toward fainter sources owing to the larger photometric errors. To 
reduce the risk of contamination by objects outside the cluster, we also impose a conservative upper magnitude limit of F160W$_{AB}$\,=\,21.0.
The initial selection flagged two objects that were found by inspection not to be
bona-fide cluster members and so were removed.  The first was a star, and the second was
galaxy ``s30"
with a spectroscopic redshift of $z$\,=\,0.135, which places it {in the foreground} 
of the cluster.  Our final catalog for G165 uses 38 cluster members in the lens model.  
The cluster members within the central region 
found by our blind selection algorithm
appear as plus signs in Figure~8.  We refer to Appendix B for additional details regarding 
the cluster member selection.

On the lensing constraints,
our \hst \ image is replete with giant arcs and arclet families.  The presence of giant arcs, as well as
structures consisting of several giant arcs, has been noted before 
\citep{Canameras:15}.  A preliminary mass model for G165 was made
using ground-based CFHT data available at the time \citep{Canameras:15, Canameras:18}. 
In total, we present here 11 designations of arclet families, all of which are new
to the literature (Table~3).
The reference center for our lensing analysis is set to the location of the lensed DSFG at 
(R.~A., decl.) = (11:27:14.731,+42:28:22.56).  By  using the positions and brightnesses of 
the cluster members as constraints in our LTM algorithm,
we construct a mass map that is subject
to the arclet family constraints.  Our model uncovers the two mass peaks evident
in the imaging and reproduces all 
lensed galaxy images 
with respect to their locations (rms $\sim$ 0$\farcs$65).  The arclet families
are marked 
on a color image along with the critical curve in Figure~8.  Postage stamp images of 
the arclet family members appear in Figure~9 organized by family name.
Below we give a description of each of the 11 arclet family systems, whose properties
are also summarized in Table~3.

{\it G165\_DSFG\_1 (Arcs 1a, 1b)}.$-${\it Arc 1a} is the NIR counterpart of the 
lensed DSFG at $z$\,=\,2.2 detected in the submillimeter data set \citep{Canameras:15}.
This giant arc, which orthogonally bridges the critical curve, has an NIR angular 
extent of $\sim$\,5$\farcs$  Our model estimates that {\it Arc 1a} is
 a merging image
with a high areal magnification factor of $\apg$\,30 that varies along the long axis 
of the arc.   The large areal extent on the sky yields the potential to study properties
within its interstellar medium, on physical scales of $\lesssim$\,1~kpc.  In particular,
two compact and bluer sources appear superimposed onto  {\it Arc 1a} which are situated on
opposite sides of the critical curve at the redshift of the DSFG.  These may potentially
be images of star forming knots within the DSFG that are multiply imaged, thereby
yielding still higher magnification factors (see Figure~5).  Spectroscopy is required
to determine the relation of these two blue images to {\it Arc 1a}.  {\it Arc 1a}
is bright, $K_{AB}= 18.74^{+0.02}_{-0.02}$ mag, and well suited
for spatially resolved spectroscopic follow-up observations. 

A counterimage is also predicted, which is not detected in our {\it HST} data set
but is detected in the LBT/LUCI\,+\,ARGOS
$K$-band image at the model-predicted location.  A bright image at the exact 
model-predicted location is also detected in our {\it Spitzer}/IRAC imaging data. 
We designate this arc as the counterimage {\it G165\_DSFG\_1b} (see Figure~5).  Interestingly, while the F160W$_{AB}$-$K_{AB}$ color is consistent between the two images, 
there is an offset in the
F160W$_{AB}$-$S_{3.6,AB}$ color by a large $\apg$2.6 mag.  This color difference 
is owing at least in part to contamination.  {\it G165\_DSFG\_1b}  appears to be situated 
behind a bluer and lower-redshift galaxy, which influences the photometry and 
therefore renders the color unreliable (see Figure~5).  {\it G165\_DSFG\_1a} is also a 
merging pair.  As such, the background source crosses a cluster caustic
such that {\it G165\_DSFG\_1a} represents only a region of that background source
and only a portion of
the starlight.  At the same time, {\it G165\_DSFG\_1b} unveils the entire source
 and thus the total integrated galaxy light.  It is noteworthy that
{\it G165\_DSFG\_1} is the only arclet family in this field to have a measured spectroscopic redshift.  This family is used for the internal minimization or 
``anchor" of our model.

{\it G165\_2a, 2b, 2c (Arcs 2a, 2b, 2c)} and {\it G165\_8a, 8b, 8c (Arcs 8a, 8b, 8c)}.$-$The {\it Arc 2} family members are the brightest in the field, with {\bf $K_{AB}$}
magnitudes for each of the three arcs of $\approx$18.5 mag, making them also excellent 
sources for follow-up spectroscopy to measure the redshift.  
For this object, we leave the redshift to be optimized in the modeling.
The bluer arclet trio that makes up  {\it Arcs 8}, which are 
situated near in projection,
are undetected at $K$.   {\it Arcs 2a} and 
{\it 2b} and {\it Arcs 8a} and {\it 8b} fold about an axis of symmetry, as do {\it Arcs 7a} 
and {\it 7b} and {\it Arcs 10a} and {\it 10b} discussed below.

 {\it G165\_3a, 3b, 3c} ({\it Arcs 3a, 3b, 3c}); {\it G165\_4a, 4b, 4c} ({\it Arcs 4a, 4b, 4c}) and {\it G165\_6a, 6b, 6c} ({\it Arcs 6a, 6b, 6c}).$-$
For the following description we refer to the close-up image in Figure~5.   
The family members {\it Arcs 6a} and {\it 6b}  are red and compact arcs that
are situated on 
opposite sides of an axis of symmetry, as marked.  Adjacent in projection on the sky, 
the slightly redder family members {\it Arcs 4a} and {\it 4b} present more extended 
morphologies.  Coincident with {\it Arcs 4a} and {\it 4b}, the bright family members 
{\it Arcs 3a} and {\it  3b} describe a fold arc conjoined at the axis point.  The third 
image of each of these families, {\it Arcs 3c, 4c}, and {\it 6c}, appears at an an angular 
separation of $\approx$14$\farcs$   This set of third images for each family retains 
similar colors and image morphologies and relative image placements.

 {\it G165\_5a, 5b (Arcs 5a, 5b)}.$-$
These faint and blue galaxy images are situated just inside the critical curve and are the only 
secure arclet family members to reside on the opposite side of the gravitational potential.  
{\it Arcs 5a} and {\it 5b} are two merging images folded about the
 critical curve.   Meanwhile, the dashed circle labeled as ``5c?" marks the position of a candidate counterimage
 that awaits confirmation pending additional model constraints.

{\it G165\_7a,7b,7c (Arcs 7a, 7b, 7c); G165\_10a,10b (Arcs 10a, 10b)}.$-${\it Arcs 7a} and {\it 7b} and 
{\it Arcs 10a} and {\it 10b} project onto an arc-like structure that is parallel 
to  {\it Arcs 2a, 2b,} and {\it 2c}.  
{\it Arcs 7a} and {\it 7b} are especially  
red and  low in surface brightness.  The counterimage that we designate as
{\it Arc 7c} appears southward  
at the model-predicted location.  The candidate counterimage labeled as ``10c?" appears near to its expected
location but at a different color, and so it is not included in our lens model.

 {\it G165\_9a, 9b, 9c (Arcs 9a, 9b, 9c)}.$-$This arclet family trio is
 distinctively blue and compact.  {\it Arcs 9a} and {\it 9b} are split by an axis of symmetry. 
 {\it  Arc 9c}
 appears at the model-predicted location at an angular separation of 10$\farcs$ 
 Note that two other candidate counterimages are marked in Figure~8 as ``9d?"
 and ``9e?" on the opposite
 side of the gravitational potential, which await confirmation
 as additional model constraints become available.
Although situated near in projection to bright central elliptical galaxies, {\it Arcs 9d? and 9e?} are clearly
 identified in our galaxy-subtracted image using {\it Galfit}  \citep{Peng:10} in Figure~9.
 
{\it G165\_11a, 11b, 11c (Arcs 11a, 11b, 11c)}.$-$The blue {\it Arcs 11a} and {\it 11b} are images that merge across the critical curve
 as indicated by the pair of star forming knots within {\it Arc 11a}
that appears again in {\it Arc 11b} with reverse parity.
{\it Arc 11c} appears at the model-predicted location southeast of the other two arclet family images
at an angular separation of 18$\farcs$

From our lens model we compute a large effective Einstein radius of 13$^{\prime \prime}$ at $z$\,=\,2.2 
and 16$^{\prime \prime}$ at $z$\,=\,9.  By integrating up the mass surface density, we measure a lensing 
mass of (2.6\,$\pm$\,0.11)\,$\times$\,$10^{14}$ $M_{\odot}$  within a $\sim$\,250 kpc radius.   By 
summing up the total area on the magnification map binned by the magnification factor, we compute 
A($>$\,$\mu$) as a function of  $\mu$.  Our profile of this cumulative areal magnification is similar to that of the 
Weak and Strong Lensing Analysis Package (WSLAP) model of \citet{Diego:07}, to within $\sim$\,30\,\%.  Note that 
given the significant visibility of both {\it G165\_DSFG\_1a} and {\it G165\_DSFG\_1b} in the $K$ band and 
{\it Spitzer}/IRAC, the James Webb Space Telescope  ({\it JWST}) resolution and sensitivity will be needed at 
1\,--\,4 $\mu$m to significantly refine these lens models.  We refer to \S~6 for independent measurements of the 
mass and estimates of the lensing strength and also to Appendix B for the magnification map. 

\section{Discussion}

\subsection{\sc The Mass of G165}

The difference in our values between the lensing and the dynamical masses merits 
further investigation.  Here we discuss our three independent 
estimates of the mass.

\subsubsection{Lensing Aass}

We measure a lensing mass of $(2.6 \pm 0.11)$\,$\times$\,$10^{14}$\,M$_{\odot}$  within $\approx$\,250 kpc by applying the constraints imposed by the 11 arclet families (\S 4.2).  Of these, we have spectroscopic confirmation only for 
{\it G165\_DSFG\_1a} of $z$=2.2357 \citep{Harrington:16}.
We choose to allow the redshifts of other arclet families 
to vary as free parameters with values of $z$\,=\,1.5\,--\,7.  While the approach works reasonably well 
in that it yields accurate 
model predictions of the counterimages, nevertheless, the lack of redshifts is nonideal.  
This is because uncertainties in the lensed galaxy redshifts 
 translate into uncertainties on the normalization of the lens model, 
 which in turn lead to changes in the value for the total mass of dark plus visible matter.
 
 We find the mass density
 to fall off rapidly beyond 250 kpc, and to reach $\approx$\,$4 \times 10^{14}$\,$M_{\odot}$ within 1 Mpc.
 This value is lower than our value for the dynamical mass by an order of magnitude.
 This then raises the question whether
an external shear component may be situated in such a way that it controls, or at least contributes
to the determination of the positions and orientations of the lensed images.
If so, then such a structure could 
potentially hide additional mass outside of the {\it HST} field of view that would
not be accounted for in our strong-lensing
mass estimate.  There are extended structures 
in our wider-field (4$^{\prime}$\,$\times$\,4$^{\prime}$) 
LBT LUCI-ARGOS $K$-band image, yet our LTM model does not 
uncover any significant external shear component.  At the same time, our model covers only the inner portion of a large and extended 
lens. 
Additional deep and wider-field imaging is needed to extend the model into the weak-lensing regime
to investigate the influence of any external lensing structures.

\subsubsection{Dynamical Mass}

Our value for the dynamical mass of 
$1.3^{+0.04}_{-0.70} \times 10^{15}$\,$M_{\odot}$
is a factor of $\sim$5 higher than that of the lensing mass within 250 kpc.  
By making use of our entire spectroscopic data set, which extends to 1 Mpc, our value for the dynamical
mass remains high, $M_V$\,= \,$(9.1 \pm 0.4)$\,$\times$\,$10^{15}$\,$M_{\odot}$.
Relevant to this discussion, the imaging uncovers an obvious bimodal mass structure  (Figure~8). If the mass is
elongated along the line-of-sight, then the velocities will also be  
higher in this direction. In this case the erroneous assumption that the
line of sight velocity is spherically symmetric will lead to an overestimate of the virial mass.
   Bimodal masses are not uncommon in massive
lensing clusters \citep[e.g.,][]{Cerny:18,Cibirka:18,Mahler:18}.  For example, in \citet{Mahler:18}, 
the two mass peaks appearing in the image of the cluster A2744 are 
identified {\it spectroscopically}
in the redshift histogram of 156 cluster members
as two velocity peaks separated by 5000 km~s$^{-1}$ (their Figure~9).  We cannot perform this exercise in our current sample, given the lower numbers of spectroscopic
 redshifts by an order of 
 magnitude.  Instead, we undertake a search for any
velocity gradient across the cluster.

 \begin{figure}[h!]
\includegraphics[width=1.0\linewidth]{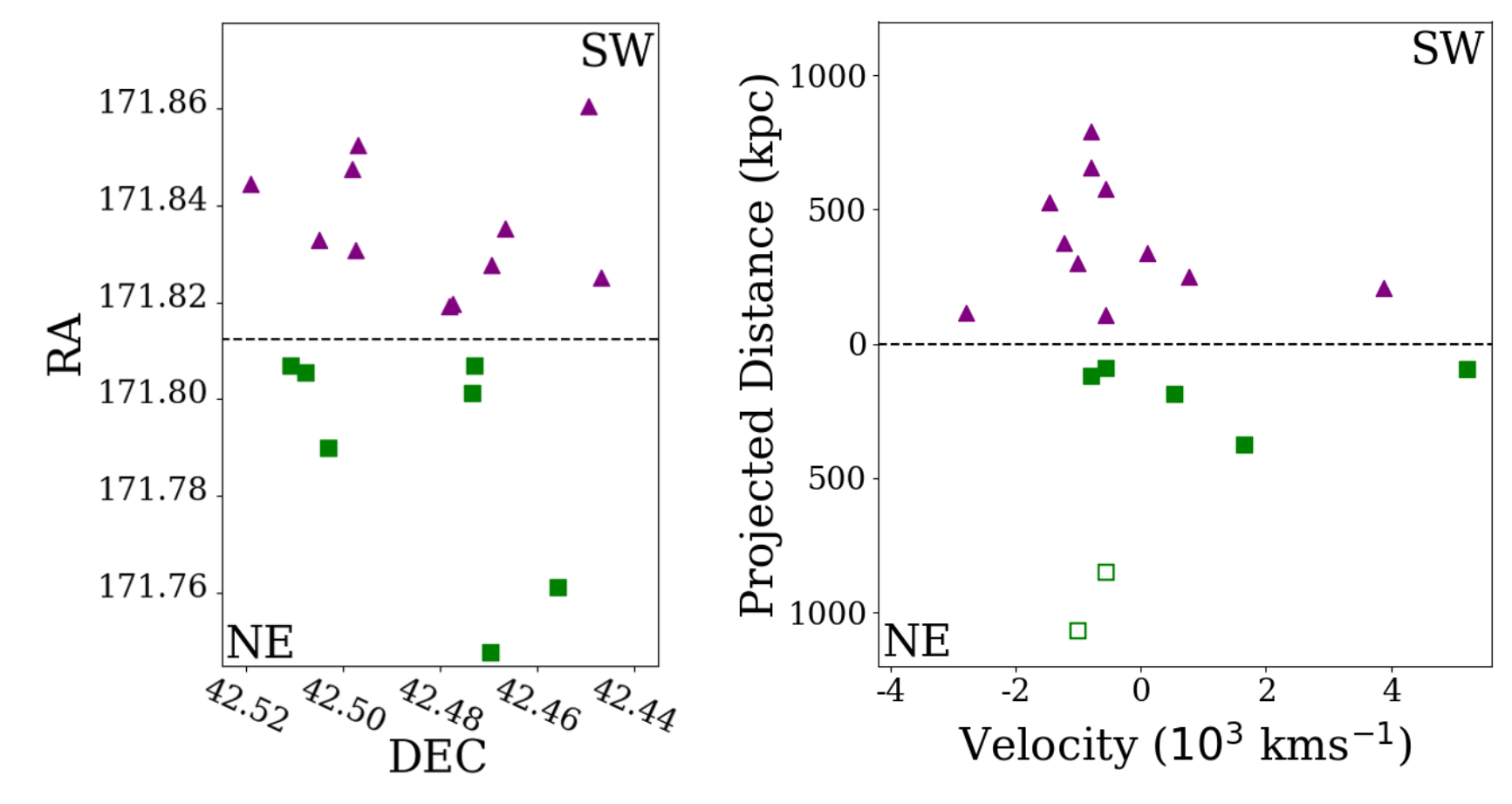}
\caption{Left: scatter plot of (R.\,A., decl.) for the cluster members in G165 with spectroscopic 
redshifts.  We divide the cluster into an NE and an SW region by a bifurcation line (dashed) drawn 
at the midpoint between the two mass peaks.  The cluster members on either side of the  line are 
indicated by the square- and triangular-shaped symbols, as marked.  Right: scatter plot of the 
projected distance from the  bifurcation line (dashed) as a function of the velocity relative to the 
cluster mean redshift of $z$\,=\,0.351.  The  two green open squares pertain to objects with high 
negative velocities and that also show relatively rare nebular emission lines in their spectra.  
If we remove these two outliers under the assumption that they may represent new infalling galaxies, 
then the remaining data may suggest a velocity gradient across the cluster.  However, using this 
data set,  we cannot place firm constraints on the lensing configuration of G165.
}
\end{figure}

We introduce a bifurcation line drawn normal to the line connecting the NE and SW 
mass peaks at its midpoint (see Figures~8 and 10).  We then compute the radial velocities 
on either side of this line to search for evidence of two velocity peaks to match the incidence
of the two mass peaks (Figure~10).  
But do the velocities give a fair representation of the kinematical 
structure of the cluster?  In a recent paper,
\citet{Hayashi:17} report that new cluster members undergoing infall 
show high line-of-sight velocities at {\it all} radii.  This is potentially insightful
for the G165 field, in which two cluster galaxies have high measured velocities
of $v_{los} \approx 950$ and 1750 km s$^{-1}$ (green open squares in Figure~10).
These are also two of only three galaxies
showing nebular emission line features  
indicative of recent star-formation.
These galaxies have
a potentially larger peculiar
velocity component and elevated star forming activities, two attributes
that are  consistent with the
picture that these objects are infalling members.
On consideration of all but these two outliers, there is a hint of a redshift of the NE mass peak relative to the 
SW one.  However, given the small mean velocity difference between the two peaks of 
$\apll$2000 km s$^{-1}$, and the dearth of spectroscopic redshifts, we are unable to constrain the cluster 
velocity configuration with the current data set.  Additional spectroscopy is needed to fill in the sparse redshift sampling  to obtain 
a larger, more representative set of cluster members.  We refer to Section~6.3 for the discussion of {\it how} the cluster
configuration relates to the cluster gas pressure.

\bigskip
\bigskip
\bigskip

\subsubsection{Caustic Mass}
In keeping with the limits typically imposed for the caustic mass estimates, we set 
a velocity cut of $\pm$\,5000 km~s$^{-1}$ 
from the mean cluster redshift of $z$\,=\,0.351.  The redshift information
for the 17 of 18 cluster members meeting this requirement provides the input to measure 
the caustic mass in a formalism developed in \citet{Diaferio:97} and \citet{Diaferio:99}
\citep[see also][]{Serra:11,Alpaslan:12, Windhorst:18}.
The approach is to estimate the mass of a cluster of galaxies out to the virial radius by analyzing the distribution of its constituent galaxies in redshift space (i.~e., ~projected separation from the cluster center $R$ as a function of line-of-sight velocity with respect to the cluster median redshift $v_{\mathrm{los}}$). 
On the assumption of a virialized cluster, this distribution resembles the bell of a trumpet (with the spread in $v_{\mathrm{los}}$ increasing at low $R$), whose area can be related to the gravitational potential (and hence mass) of the cluster.

It is useful  to work in phase space by 
depicting $v_{\mathrm{los}}$  as a function of their
projected distances from the cluster center.  
We adopt the virialized region from the
prescription in \citet{Jaffe:15}, such that $v_{los}$\,$\leq$\,1.5\,$\sigma$ is within 
a projected distance of $R_{200}$, where $\sigma$ is
the velocity dispersion. 
Indeed, the vast majority of cluster members (black circles in Figure~11) fit well within this  radius as depicted by the gray shaded region. 
We convert our redshift catalog of cluster members 
into a continuous density field by using an adaptive density kernel.
The contour (black curve) identifies the region in the redshift-space distribution that corresponds to the escape velocity of the cluster (assuming spherical infall), which in turn is related to its gravitational potential as $v^2_{\mathrm{esc}} = -2\phi(r)$. In practice, we impose the condition of spherical symmetry 
by rewriting this  density threshold into a symmetric version about the $v_{\mathrm{los}} = 0$ line.
To do this, we check the absolute values of $v_{los}$ for this double-valued function 
in small increments of radius along the density threshold 
contour.  The caustic equates to
 the minimum of those two absolute values and is reflected along the $v_{los}$\,=\,0 line to construct
 the ``tuning fork" shape (green contour).  The amplitude of the caustic $\mathcal{A}(r)$ is then related to the cluster mass $M$ such that $GM \propto \int_0^r \mathcal{A}^2(r)dr$.  
 
  \begin{figure}[h]
\includegraphics[viewport=5 0 200 480,scale=0.48]{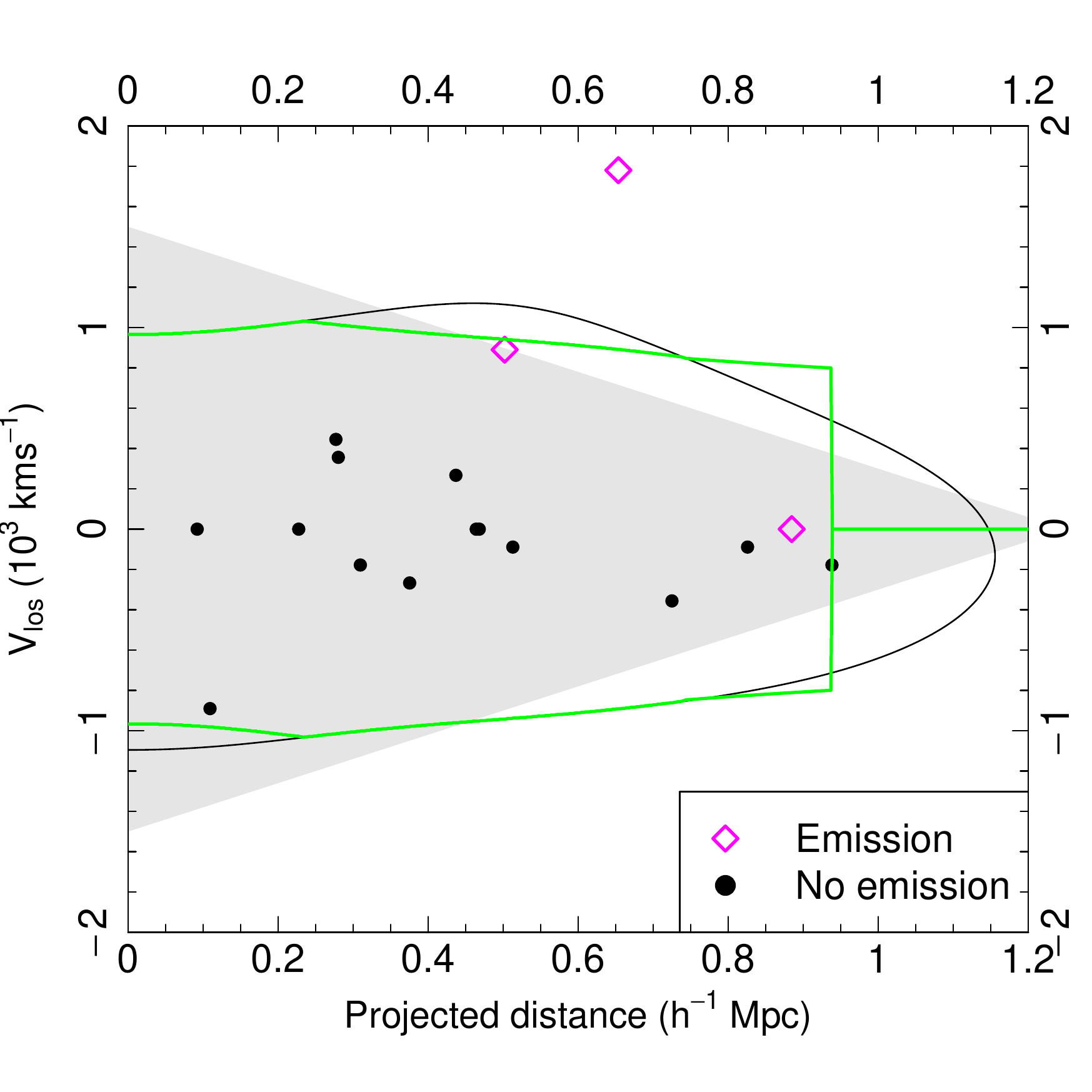}
\caption{``Trumpet" diagram depicting the caustic mass estimator for G165.  We start with
the redshift
catalog of cluster members (black circles), which we convert into a density field.  The contour (black) 
identifies the region in redshift space at which the galaxy density equates to some threshold value, 
which in this case corresponds to the
escape velocity of the cluster.  We rewrite the equation for this threshold density 
into  a form that is symmetric about
$v_{los} = 0$ (green ``tuning fork") to satisfy the requirement of spherical symmetry in our model.
The mass estimate  is then made by integrating the area
under the caustic lines.  We measure a mass of 
$M_{caustic}$\,=\,(1.9\,$\pm$\,0.18)\,$\times$\,$10^{15}$\,M$_{\odot}$
within $r\approx0.8$~Mpc.}
\end{figure}

\begin{figure*}
\includegraphics[width=1.0\linewidth]{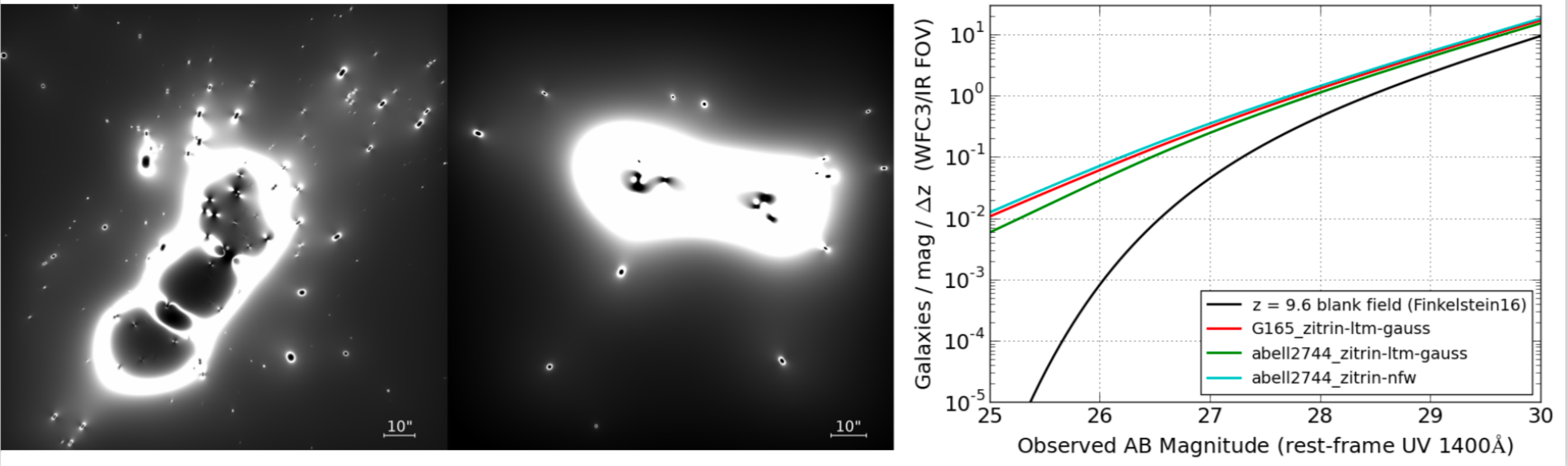}
\caption{Comparison of the G165 lens {\bf ($z$=0.351)} with HFF cluster 
A2744 ($z$=0.308).  The two left panels show magnification maps 
corresponding to the $z$ = 9 critical curves for A2744 (left) and G165 
(middle) obtained from our LTM methodology \citep{Zitrin:15}.  The right 
panel shows a {\bf $z$~=~9.6} luminosity function for a blank field (black) 
placed in the background of our LTM model for G165 (red), our LTM 
model for A2744 (green), and a fully parametric model  (blue).  The two clusters yield similar numbers of lensed
high-redshift galaxies for similar exposure times.  Our lens model suggests that G165 has a similar 
lensing strength to HFF cluster A2744.  Note that the LTM methodology tends toward a
flatter mass distribution compared to fully analytic techniques, as discussed in
\citet{Zitrin:15}.}
\end{figure*}

By applying this
estimator, we measure a mass of $\approx$$(1.9 \pm 0.18)$\,$\times$\,$10^{15}$\,M$_{\odot}$ within 0.8 Mpc.  
The uncertainty on this value is derived by a ``jackknife" resampling approach 
consisting of making
20 realizations in which two galaxies at a time are removed at random and the
mass recomputed.  Analysis of this set yields the stated estimate in the uncertainty
of the mass. 
Note that the mass has been 
rescaled to be median biased with respect to the dynamical mass, which is
calibrated as a function of redshift and cluster richness of comparable systems in \citet{Alpaslan:12}.  
This mass value is a factor of $\sim$\,5 higher than the value for the lensing mass extrapolated
out to 1\,Mpc, and a factor of  $\sim$5 lower than 
the value for the dynamical
mass computed within 1 Mpc. 
 If G165 does have an aspherical mass distribution elongated
along the line of sight (see \S 6.1), then  this value will be an overestimate.

\subsection{\sc G165 as a Lens}

We compare  the lensing strength of G165 with that of another massive lensing cluster 
at a similar lens redshift, the Hubble Frontier Fields cluster A2744  (HFFs; PI:  J.~Lotz,
GO-13495).  A2744 provides a useful benchmark for its well constrained lens model 
and similar size of its effective Einstein radius.  Its strong-lensing model is 
well constrained with 29 arclet families identified  from deep \hst \ imaging
in seven bands with 5$\sigma$ limits in each filter of $m_{AB}$\,$\sim$\,29 mag 
\citep{Mahler:18}.  These limits are  $\sim$1.3 and $\sim$1.8 AB mag deeper than the 5$\sigma$ limiting magnitudes
for G165 of F110W$_{AB}$ = 27.7 mag and F160W$_{AB}$ = 27.0 mag.  For consistency, we construct the models for both clusters by our LTM 
approach, where the lens model for G165 comes from this paper and the one for
A2744 is from \citet{Zitrin:14}.   We show the lens models in {\bf the}
two leftmost panels in Figure~12.  In both cases
we find a similar elongated shape and effective $z$\,=\,9 critical 
curve size of $\sim$16$\farcs$
To compute the lensing strengths, we assume the same background luminosity function 
 \citep{Finkelstein:16}, and then compare the number distribution of lensed background 
 galaxies in the two fields.   Overall, the clusters G165 and A2744 yield significantly brighter 
 objects compared to a blank field at all magnitudes.   At high redshifts, the clusters G165 and A2744
yield on average similar numbers of {\bf $z\approx 9.6$} objects (right panel in Figure~12).

G165 is an ideal lens through which to investigate high-redshift objects 
($z$\,=\,9\,--\,12).  This is in part due to the relatively low redshift of the 
lens plane of $z$\,=\,0.351, for which the level of the intracluster light 
(ICL) contamination at the NIR wavelengths corresponding to the Lyman break 
for $z$\,$>$\,9 galaxies is minimized \citep{Windhorst:18}.  G165 also has a 
reasonably high ecliptic latitude of 35$^{\circ}$, reducing its background from 
the peak with the zodiacal plane. The lens size is ideal for JWST/NIRCam 
imaging, as the lens fills (but does not overfill) the field of view out to 
$\sim$\,3 times its Einstein radius.  Note that for relatively shallow exposures 
typical of a JWST short program reaching limiting fluxes of 
$m_{AB}$\,$\sim$\,28 mag.  
For such cases, the improvement of JWST will come 
at all wavelengths, and especially from imaging at $\lambda$\,$>$\,1.6$\mu$m 
which enable  robust detections of the stellar continuum of any new high-redshift 
galaxy candidates situated behind lensing clusters.

We consider the potential for {\it G165\_DSFG\_1a} to yield caustic transit 
events.  In its favor, two compact and bluer knots appear in projection to be within this arc that bridges
the critical curve and have small angular separations from the critical curve of roughly tenths of an arcsecond
 (see Figures~5 \& 9).   
Unlike distributed masses that incur magnification factors of up to $\mu$\,=\,40\,-\,50, 
compact sources such as stars and star clusters can reach magnifications of $\mu$\,$\apll$\,$10^5$ \citep{Windhorst:18}. 
To assess the practicality of monitoring  this  arc  for  caustic  transits,  we require 
in addition a  negligible  ICL  component. \cite{Diego:18} showed how, if the ICL at the position of the critical curve is significant, the stars from the ICL can set an upper limit (through microlensing) on the maximum magnification of background stars during caustic crossing events to around $10^4$. Since the ICL can extend up to large distances from the center of the halos \citep[see e.g., ][]{Mihos:16}, even critical curves that are relatively far from the center of halos could be affected by the microlensing from the ICL stars.

\subsection{\sc The G165 Cluster Gas Pressure}
Given the somewhat novel search strategy to find the G165, it is natural to ask how this massive 
lensing cluster compares with others selected by more commonly used methods, such as 
X-ray brightness or the  detection of the SZ decrement.  G165 has high mass and high dark matter 
concentration, as evidenced by the  prominent displays of giant arcs and arclet families even in 
these relatively shallow (single-orbit) \hst \ images.  As such, we would expect for G165 to be replete with large 
amounts of cluster gas.

G165 is in fact undetected in ROSAT imaging (R6+R7 bands, or $\sim$\,0.7\,--\,2 keV).  
Put another way, G165 is at best a low-luminosity X-ray source with an upper limit 
on its X-ray flux computed from the RASS diffuse map of  $1.12 \times 10^{-4}$ 
counts s$^{-1}$ arcmin$^{-2}$.   It is unusual for a truly relaxed cluster to have an 
X-ray flux so low as to be undetected by ROSAT at this redshift and mass scale.  
At the same time, at these lower luminosities, the scaling relations correlating the X-ray 
luminosity to cluster mass are more uncertain owing to a large intrinsic scatter in the 
data  \citep{Bruch:10}. G165 also misses out on membership in the {\it Planck} 
Sunyaev-Zel'dovich (PSZ) cluster catalog as a result of its low SZ signal, which 
falls below the minimum detection threshold. In the {\it Planck} Compton-{\it Y} parameter 
map there is a small fluctuation at the position of the cluster that may represent a
weak detection of intercluster gas, or it may be noise given that the detection is only at the 
1$\sigma$-2$\sigma$ level.  This lack of a significant SZ signal might be  a consequence of radio 
emission washing out a shallow decrement, projection effects, or an overestimation of the 
cluster mass.

Radio sources have an inverted spectrum with respect to IR sources that can counteract 
the SZ signal.  As {\it  DSFG\_G165\_1a} is the one image in the field with high submillimeter 
flux arising from high  star-formation and/or AGN activity; this lensed DSFG is the most likely 
source to be radio-loud.   There is a weak radio emitter detected near to the position of the 
IR source.  From NVSS data we measure a total flux from the cluster including this IR source 
of $<$\,40 mJy at 1.4 GHz \citep{Condon:95}. Although present, this modest radio signal is 
insufficient to compete with the SZ effect at the relevant frequencies (100\,--\,353 GHz), thereby 
ruling out radio contamination as an explanation for the relative SZ silence.

The last conventional explanation is the lensing configuration. The G165 field contains an 
obvious bimodal substructure.  There are other examples of bimodal mass structures,
 such as the well-studied Bullet Cluster \citep{Bradac:06, Clowe:06}, 
and the cluster known as ``El Gordo" \citep{Menanteau:12}.  These two clusters are classified as ``post-mergers"
that, in turn, produce significant enhancements of the X-ray flux.
If the field is elongated along the line-of-sight 
direction as a series of two smaller galaxy structures, then we may be catching G165 during a 
less well studied evolutionary ``pre-merger" phase.  In this scenario, the total cluster gas 
pressure dilutes across the large structure, which reduces the gas pressure 
and the X-ray emission, hence reducing the SZ decrement. 
At the same time,  the mass integrated along the line of sight still provides ample
surface mass densities, leading to 
 strong-lensing effects.  A test of this hypothesis can be made by acquiring
 X-ray observations. 
 For example, {\it XMM} imaging at the level 
 of 20$-$27 ks total exposure allows us to measure the distribution and 
 centroid and place
 limits on the electron temperature of the X-ray gas.  If G165 deviates from a monolithic
 mass, then an offset will be detected, or at least an X-ray source that is 
 marginally extended yet still potentially offset from the center of mass.

\section{\sc Summary}

Searching wide-field imaging data sets for giant arcs is now fairly common, yet
conducting searches for unresolved giant arcs at submillimeter wavelengths is still 
relatively rare.  We obtained \hst \ WFC3-IR imaging of the fields of six lensed DSFGs
selected in a novel search by their rest-frame FIR color and compactness using
{\it Planck/Herschel} data.  We extend the analysis on the G165 field, which 
shows prominent giant arcs and arclet families. We find the following:

\begin{enumerate}

\item {Each of our six sample fields unveils the  {\it HST} WFC3-IR counterpart of the strongly lensed DSFG.  
In four fields, the DSFG image appears as an image multiplicity {\it HST} resolution (G165, G045, G145, and G080).}

\item For the G165 field, we obtained ground-based spectroscopy using MMT/Hectospec and 
{\it Gemini}/GMOS.  We measure 51 new redshifts, which augment the spectroscopic
catalog of objects in this field by a factor of five.  For the five cluster members 
within 250~kpc, we compute
a velocity dispersion and then apply the virial theorem to estimate the dynamical mass of $1.3^{+0.04}_{-0.70} \times 10^{15}$\,$M_{\odot}$. Using our full catalog of 18 cluster members,
 we compute a dynamical mass of $M_V$\,=\,(9.1\,$\pm$\,0.4)\,$\times$\,$10^{15}$\,$M_{\odot}$ within 
 1~Mpc.
We also estimate
the caustic mass, which is (1.9\,$\pm$\,0.18)\,$\times$\,$10^{15}$\,$M_{\odot}$ within
$\sim$\,0.8 Mpc.  These mass estimates are high, possibly owing to enhanced velocity
structure in the line-of-sight direction and/or several nonvirialized cluster substructures
adding to the lensing power.

\item{ For the G165 field, we acquired LBT LUCI+ARGOS $K$-band 
imaging at high resolution (FWHM\,$\approx$\,0$\farcs3$).  The $K$-band image
uncovers dozens of lensed galaxies,
 including 11 arcs drawn from five different arclet families.
We confirm the image position of the lensed DSFG, {\it Arc~1a}.
We also 
make the first detection of its counterpart, {\it Arc 1b}, at the model-predicted location, which is
is too faint and too red to detect in our {\it HST} data set.}
 
\item In total, for the G165 field we identify 11  arclet families by their similar colors and morphologies,
which are all new.  Obvious axes of symmetry  corroborate our arclet family designations.
In the NIR, {\it Arc~1a} subtends 5$^{\prime \prime}$ and is magnified by a factor of  $\apg$30. 
{\it Arc~1b} is fainter and detected only in our high-resolution LBT/LUCI\,+\,ARGOS
$K$-band image and {\it Spitzer/IRAC} images.  We measure an
F160W$_{AB}$-$S_{3.6}$ color difference between the two images that arises because  {\it Arc~1a} is 
a merging image and so  represents only a portion of that background source, while
 {\it Arc~1b} uncovers the entire  source image.
 
\item We use the LTM approach to construct a mass map in the fields
for which there is at least one arclet family detected in our data set
(G165, G045, G145, and G080).   
For the cases without arclet families,
we generate a $\kappa$-map through the galaxy brightnesses and orientations.
For G165, we estimate a lensing mass of $(2.6 \pm 0.11)$\,$\times$\,$10^{14}$\,$M_{\odot}$ 
within $\sim$\,250 kpc.  We compute effective Einstein radii for G165 of 
$\sim$\,13$^{\prime \prime}$
at $z$\,=\,2.2 and $\sim$\,16$^{\prime \prime}$ at $z$\,=\,9.

\item The lensing properties for G165 are not far different from those of other well-studied massive lensing clusters.
 In a counting  simulation, for G165 we predict comparable numbers of  
 high-redshift objects 
to $z$\,$\sim$\,9.6 to those of
A2744, another well-studied lensing cluster with similar lens redshift and dark matter 
properties. 

\item 
Based on the 18 spectroscopic redshifts of cluster members in G165, 
we currently lack the number of redshifts to distinguish convincingly bimodality in the velocity distribution.
Confirmation
of a line-of-sight configuration 
is impactful because it can help to explain the weak X-ray and SZ effect detections.
This is because a line-of-sight orientation will dilute 
the intercluster gas below the ROSAT and PSZ effect
detection limits, while maintaining a high surface mass density integrated
over the line of sight that amply suffices to explain the observed strong-lensing effects. 

\end{enumerate}

\acknowledgments
We appreciate helpful discussions with Eiichi Egami, Xiaohui Fan,  Dan Marrone, 
Ann Zabludoff, and Scott Tremaine, and we especially thank the anonymous referee
for comments that significantly improved this paper.   Support for program HST  GO-14223 was 
provided  by NASA through a grant from the Space Telescope Science Institute,
which is operated by the Association of Universities for Research in Astronomy, Inc., 
under NASA contract NAS5-26555.  J.M.D.~acknowledges the support of project 
AYA2015-64508-P (MINECO/FEDER, UE).  R.A.W.~was funded by NASA JWST 
Interdisciplinary Scientist grants NAG5-12460, NNX14AN10G, and 80GNSSC18K0200 
from NASA Goddard Space Flight Center.  M.\,P. and A.\,B. were funded by a UA/NASA Space
Grant for Undergraduate Research. This work is based in part on observations 
made with the {\it Spitzer Space Telescope}, which is operated by the Jet Propulsion 
Laboratory, California Institute of Technology, under a contract with NASA.
This work makes use of the Large Binocular Telescope, which is an international 
collaboration among institutions in the United States, Italy, and Germany.  We would 
like to thank the staff at {\it Gemini}-North and at the MMT
for performing the observations in service mode.  

\appendix

\section{\sc NIR Counterparts of Our Lensed DSFG Sample}

We searched for  the NIR counterparts of
the DSFG submillimeter sources.  
 Using the submillimeter positions as a guide, we detect red and relatively bright NIR counterparts for all six lensed DSFGs 
at the expected locations with respect to their positions in the submillimeter data 
\citep[][their Figure~2]{Canameras:15}.  
In all cases, the lensed DSFG images in the \hst \ images stand out as the reddest sources
in the field.  
Note that these galaxy images are significantly magnified, even if their size is smaller than or equal to the instrumental resolution of \hst.  Despite their small angular extents in some
cases, these lensed sources are still amongst the brightest DSFGs in the sky in 
the NIR owing
to their large estimated magnification factors.  

\begin{figure}[h!]
\includegraphics[width=1.0\linewidth]{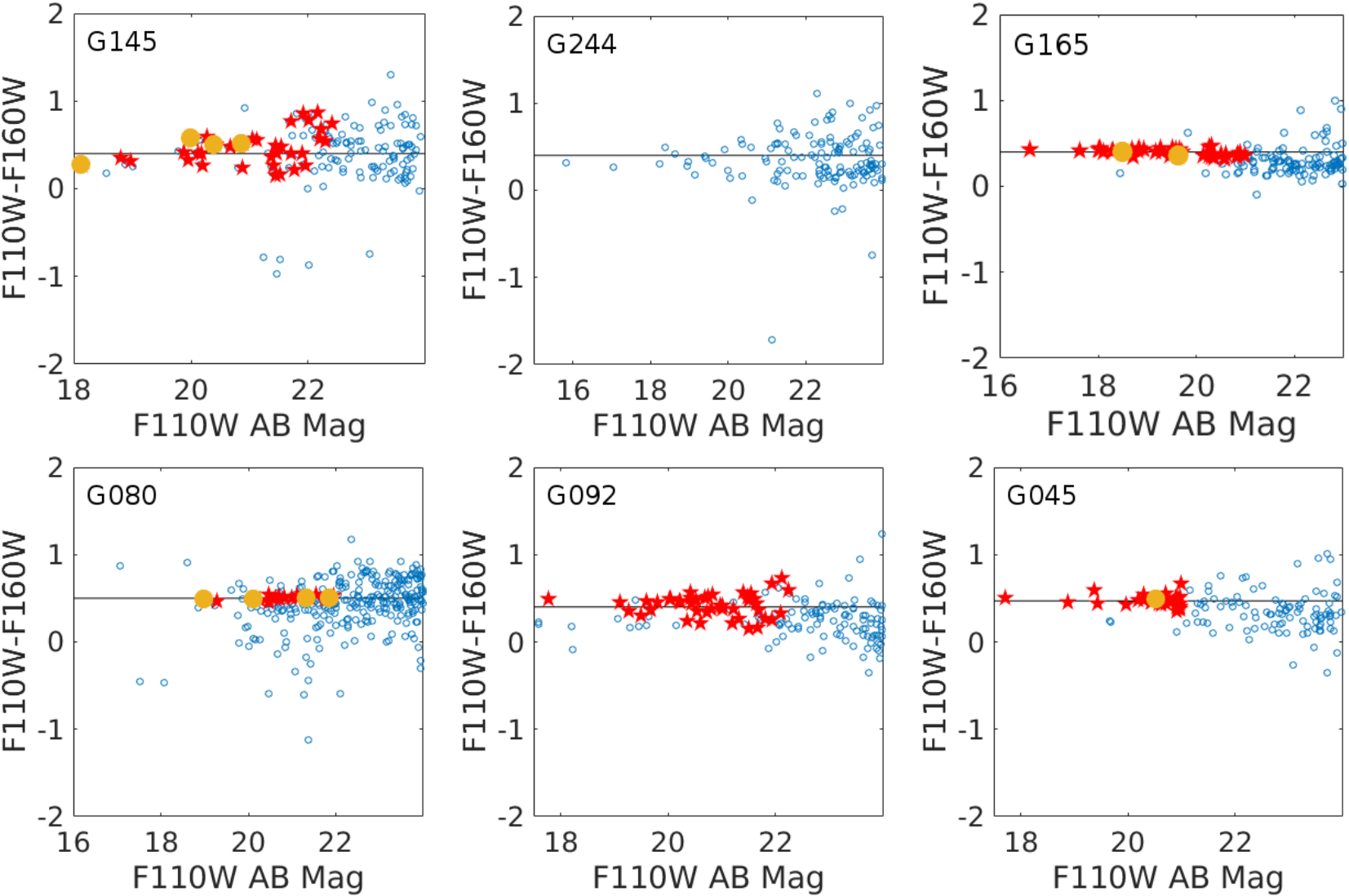}
\caption{CMDs in the six sample fields computed from our 
 \hst/WFC3-IR photometry (blue open circles).  The cluster members used in 
our lens models are indicated by the red filled stars, and the objects  
with measured spectroscopic redshifts in each cluster that are also situated in our
\hst \ images are depicted by the gold filled circles. A fiducial line 
centered on the red sequence of each lensing field is shown for reference (black 
solid line).
The G165 field shows a tight 
red sequence despite the obvious bimodal distribution of the mass, possibly indicating 
that these early-type galaxies share a similar star formation history. }
\end{figure}

In the G145, G165, G045, and G080 fields, we detect multiple images of a single background DSFG.     For G145 and G080, we find that two of the images match
up with peaks in the submillimeter (the plus signs in
Figure~2).
  In another field, G092,  the NIR counterparts are also identified but
  show different morphologies despite their similar colors.  These 
  images are more likely to be two unrelated and possibly interacting DSFGs at a similar redshift (see \S~B.5).
  For  G244, we detect the submillimeter arc but do not spatially resolve the 
Einstein ring structure, although two sets of arclet families are identified in this
field using high-resolution ALMA data  
 \citep{Canameras:17a, Canameras:17b}.
Finally, for G165 we find that
{\it G165\_DSFG\_1a} bridges the critical
curve.  
We detect another red source at the model-predicted location of the counterimage, 
 {\it G165\_DSFG\_1b}, that is prominent in our LBT/LUCI\,+\,ARGOS $K$ image and in {\it Spitzer/IRAC}  (dashed circle
in Figure~5).  The colors between the two images are different, which was initially
unexpected, as lensing is achromatic.  
At the same time, 
{\it G165\_DSFG\_1a} is an arc that is merging with an image of itself.  
Here, the background source is crossing a cluster caustic,
such that {\it G165\_DSFG\_1a} represents only a portion of that background source, while {\it G165\_DSFG\_1b} shows the entire source (see \S5.2 for more details).
The estimation of the strong-lensing properties appears below.

\section{Lensing Analysis}

We apply our well-tested LTM pipeline to the G045, G145, G092, G080, 
and G244 fields, and we refer the reader to \S 5.2 for the details 
concerning the lens model for G165.  For each field, the red lensing 
galaxies populate a distinctive region of the CMD in Figure~13.  
Galaxies on this ``red sequence" have a similar color because they 
have a similar redshift and share a  similar star formation history.  
This NIR color captures the ellipticals on the slowly varying part of 
the observed-frame SED of a several-gigayear-old elliptical galaxy, 
such that by applying a blind color cut to the {\it HST} F110W and 
F160W data, the red sequence is easily established in each field
(red filled stars in Figure~13).  To reduce the chances for 
contamination from foreground/background objects, we impose a 
conservative magnitude cut in the range of 
F110W$_{AB}$ = 21\,--\,22 mag, depending on the field.  We have 
spectroscopy within the \hst \  field of view for four clusters, G165, 
G045, G145, and G080, which aids further in their identification 
(gold filled circles in Figure~13).  Using as inputs the positions 
and brightnesses of the cluster members and the positions and 
orientations of the arclet families, we construct the strong-lensing 
model for each field.  We emphasize that all arclet families discovered, 
which include the lensing fields G165, G145, G045, and G080, are 
supported by our LTM model.  The resulting 2D magnification maps 
are plotted as ratios of the local surface mass density over the critical 
surface mass density, or $\kappa$-map, in Figure~14.  We refer to 
Table~2 for a summary of the redshifts of the lenses, the lensed 
DSFGs, and other relevant information.

\subsection{G145}
The positional centroids from the
submillimeter image are indicated by the gold plus signs in Figure~2.
We find NIR counterparts for two of these three peaks, which we designate as {\it G145\_DSFG\_1a} and {\it G145\_DSFG\_1b}.  
These two small arcs are only marginally resolved using \hst.  
Initially, only one counterpart image was identified, {\it DSFG\_G145\_1a}.
A careful search unveiled a second image with a similar color,
at the model-predicted location, which we designate as  {\it DSFG\_G145\_1b}.
Using these two arcs as inputs, the model predicts a third image that coincides with the image in the  submillimeter but which is not detected 
by {\it HST}.  The lack of a detection is not surprising, given
the faintness of the other two NIR counterparts, which both hover around the limiting magnitude
of our
observations.
The redshift distribution of galaxies in this field is broad, with a somewhat poorly defined 
peak at $z$\,$\approx$\,0.837,
which we take to be the lens plane.  This value is based on four redshifts in the 3-$\sigma$ clipped
range 0.822\,$<$\,$z$\,$<$\,0.852 drawn from our spectroscopy, 
which all fall within the \hst \ field of view (gold filled circles in Figure~13).  This spectroscopy
will appear in a separate paper (Frye et al.~2018b, in preparation).   
We note that there is {\it no} spectroscopic information available from data archives or other sources. The redshifts for the four
lensing members that are situated within our {\it HST} field of view 
appear as the
gold filled circles in Figure~13. 
Our lens model recovers both the image positions and angular separations of the counterimages with an RMS of $\sim$\,0$\farcs$1.  In turn, we estimate magnification factors of 12\,$\pm$\,0.5 and 
5\,$\pm$\,0.5
for {\it G145\_DSFG\_1a} and {\it G145\_DSFG\_1b}, respectively.
We estimate the uncertainty  by sampling the values for the magnification 
in a  neighboring annular region of width 2$\farcc$ an approach that
works reasonably well for images that are not very near in projection to  the critical curve
($\apg$\,few arcseconds).
Our model yields effective Einstein radii of 10$^{\prime \prime}$ at the redshift of the lensed DSFG and 9$^{\prime \prime}$ at
$z$\,=\,9.

\begin{figure}[h!]
  \includegraphics[width=1.0\linewidth]{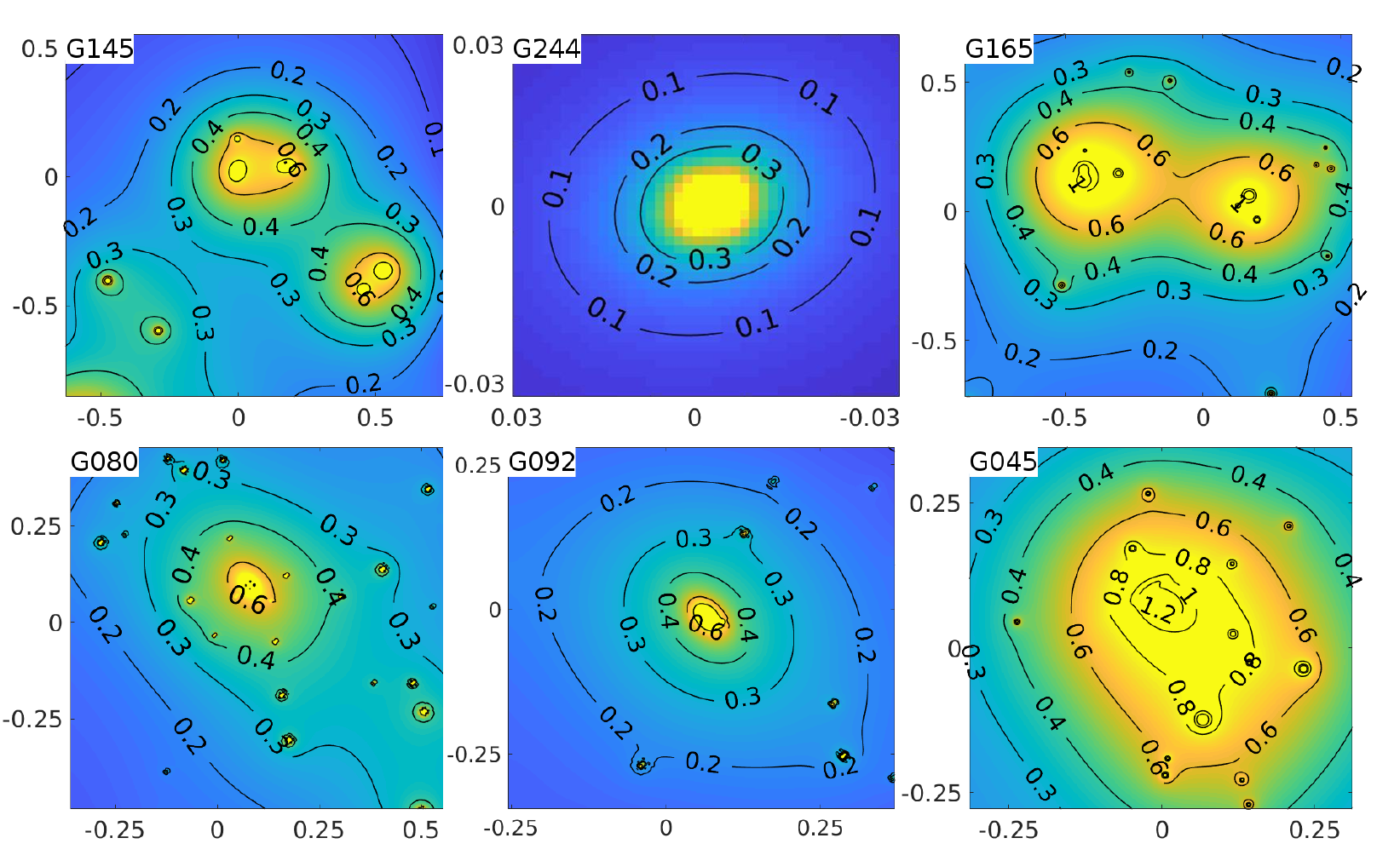}
\caption{Surface mass density distributions of the ratio of the local surface mass density to the critical value, 
or $\kappa$, for each of our sample fields using the LTM methodology \citep{Zitrin:15}.  For 
visualization purposes, $\kappa$ is scaled to the redshift of the lensed DSFG in each field.
The values for $\kappa$ are overlaid onto the contours, and the cluster orientations are the same as in Figure~2.  The reference center is given as (0,0) in each panel, and the axes increment in units of arcminutes.
The reference center for each panel
has the following (R.\,A., decl.) value in J2000 coordinates:   G145 (10:53:22.249,\,+60:51:43.93), 
G244 (10:53:53.107,\,+05:56:18.44), G165  (11:27:14.731,+42:28:22.56), G045 (15:02:36.012,\,+29:20:50.51), G080 (15:44:33.202,\,+50:23:43.53), and G092 (16:09:17.842,\,$+$60:45:19.41).}
\end{figure}
\subsection{G244}

We confirm the NIR counterpart of the lensed DSFG as a red and spatially extended image, 
although the ring-like structure and two arclet families seen in the ALMA data are blended 
with the primary lens  in our \hst \ image and are thus unresolved (Figure~2).  The primary lensing 
galaxy consists of a single object with a measured redshift of $z=1.5$,  which also  blended with the 
lensed DSFG.  Elsewhere in the field there are two blue arcs in the near projected proximity of the 
brightest cluster galaxy that appear to be unrelated images, and no other arclet families are identified. 
Without arclet families we cannot construct a lens model for this field.   Based on this information,
we are able to approximate the
surface mass density relative to the critical value through the relative galaxy brightnesses and orientations to yield
a $\kappa$-map
(Figure~14), yet we do not cite a $z$\,=\,9 critical curve radius.    
Note that this field already has a published model based on the
ALMA data \citep{Canameras:17a,Canameras:17b}.

\subsection{G045}

Four peaks of the lensed DSFG are detected in the submillimeter and ALMA imaging
\citep{Canameras:15, Nesvadba:16}.  Of these, we find NIR counterparts for three images, 
which we designate here as {\it G045\_DSFG\_1a, 1b}, and {\it 1c} (see Figure~2).   We 
measure a spectroscopic redshift for the lens of $z$\,=\,0.556, based on seven redshifts 
in the 3$\sigma$ clipped range 0.535\,$<$\,$z$\,$<$\,0.577 drawn from our spectroscopy, which will 
appear in a separate paper (Frye et al.~2018b, in preparation).  Of these, the redshift for one
cluster member is situated within the  field of view of our \hst \ data (gold filled circles in 
Figure~13).  
Our lens model recovers both the image positions 
and angular separations of the counterimages with an rms\,$\sim$\,0$\farcs$4.  In turn, the model 
yields high magnification factors of $\apg$9, $\apg$9,  and $\apg$7 for {\it G045\_DSFG\_1a, 1b,} 
and {\it 1c}, respectively.  In an independent analysis, the magnification factors of 10\,-\,22 were 
measured for smaller emissio- line regions within each arc \citep{Nesvadba:16}.    
We compute effective Einstein radii of 8$^{\prime \prime}$ at the lensed DSFG redshift and 
10$^{\prime \prime}$ at $z=9$.

\subsection{G080}
The submillimeter imaging shows three bright peaks of this one lensed DSFG.
The positional centroids of the peaks are indicated in Figure~2 by the gold 
plus signs and labels.  We designate the two NIR counterparts that we detect in our 
\hst \ imaging as {\it G080\_DSFG\_1a} and {\it G080\_DSFG\_1b}.  There is
considerable noise at the expected positions of the images owing to 
their close projected proximity to the extended halos of bright lensing galaxy members.     
We found that by smoothing the data we were able to take out the high-contrast noise, 
an exercise that enabled the detection of  {\it G080\_DSFG\_1a} and {\it G080\_DSFG\_1b} 
(see inset of Figure~2).  Interestingly, we measure a shift by up to {\bf 0$\farcs$5} in the 
positional centroids of {\it G080\_DSFG\_1a} and {\it G080\_DSFG\_1b} between the 
{\it SMA} and {\it HST} images, equating to a physical extent in the source plane of 
$\sim$\,4 kpc.  We find no good explanation for these positional offsets.  We measure a 
lens redshift of $z$\,=\,0.670 that is based on 10 redshifts in the 3$\sigma$ clipped
range $0.649 < z < 0.691$ drawn from our spectroscopy in this field, whose results will 
appear in an upcoming paper (Frye et al.~2018, in preparation).  Of these, the redshifts of 
four of the cluster members are situated within the field of view of our \hst \ data (gold filled 
circles in Figure~13). In general, the red sequence shows somewhat 
more scatter than in some 
of the other fields, which introduces a higher probability for misidentifying objects of roughly 
similar colors.  To mitigate any potential contamination by galaxies external to the cluster,
we make a conservative color cut, resulting in a narrow band of cluster members
(red filled stars in Figure~13), yet 
the uncertainty on the placement of this narrow color cut ultimately
limits its usefulness.
 Our lens model recovers both the image positions and 
angular separations of the counterimages with an rms $\sim$\,2$
\farcs2$.   The relatively 
low value of the rms uncertainty shows that the model is robust.  At the same 
time, the rms value is higher than those computed for the other fields in our sample
for two reasons:  (1) there is a higher uncertainty on the positional centroids of  
 {\it G080\_DSFG\_1a} and {\it G080\_DSFG\_1b} given their ultralow surface brightness,
and (2) the high scatter in the red sequence translates into a higher probability for 
contamination by objects with similar colors that are not bona fide lensing galaxy members.
From our lens model high magnification factors of $\sim$\,20 are measured for each of the two 
images.  We compute effective Einstein radii of $\sim$\,7$^{\prime \prime}$
at the redshift 
of the lensed DSFG and $\sim$\,$11^{\prime \prime}$ at $z$\,=\,9.

\subsection{G092}
The single ``tadpole-shaped" arc detected in the SMA imaging breaks up into two lensed 
sources, {\it G092\_DSFG\_1a} and {\it G092\_DSFG\_1b}, in our \hst \ images.  These arcs 
are not easily reproduced by our lens model despite their similar colors.  A clue to their nature 
is given by subtracting off the light of the central elliptical galaxy using {\it Galfit}.  By doing this, 
we uncover significant differences in the smooth vs.~clumpy components of the two images 
(Figure~2, inset).  The measured redshift is $z$\,=\,3.3, which is integrated over both components.  
Based on the available information, we infer that these two images are two different galaxies at 
a similar redshift.  As such, this may potentially be an example of a pair of interacting galaxies 
that induces the ultrahigh star-formation rates of $\sim$\,1000 $M_{\odot}$ yr$^{-1}$ obtained 
from correcting the value in \citet[][their Table~2]{Canameras:15} by the magnification factor 
provided from our lens model.  There is only a single available redshift in this field from the 
literature, which is of high value, as it corresponds to that of the central lensing galaxy 
($z$\,=\,0.448 from SDSS DR~14).
Without an arclet family, we cannot construct a robust lens model for this field. 
At the same time, we are able to approximate the
surface mass density relative to the critical value through the relative galaxy brightnesses and orientations to yield
a $\kappa$-map
(Figure~{\bf 14}). 
By adopting our best-fit scenario
 that {\it G092\_DSFG\_1a} and {\it G092\_DSFG\_1b}
 are two singly imaged lensed sources at a similar redshift, we compute
  high magnification factors of $\sim$20 for each image, and we do not cite a  $z$\,=\,9 critical curve radius.

\end{document}